\documentclass[10pt,journal,compsoc]{IEEEtran}

\usepackage{xcolor}
\usepackage{soul}
\usepackage{cite}
\usepackage{multirow}
\usepackage{amsmath}
\usepackage[bookmarks]{hyperref}
\usepackage{algpseudocode}
\usepackage{numprint}
\usepackage[binary-units,per-mode=symbol]{siunitx}
\usepackage{mathtools}
\usepackage{multirow}
\usepackage{lipsum} 
\usepackage{enumitem}
\usepackage{booktabs}
\usepackage{lipsum}
\usepackage{verbatim}
\usepackage{caption}
\usepackage{float}
\usepackage{tikz}
\usepackage{amsfonts}
\usepackage{diagbox}

\renewcommand{\eqref}[1]{(\ref{#1})}
\newcommand{\secref}[1]{\mbox{Section~\ref{#1}}}

\newcommand{\figref}[1]{\mbox{Figure~\ref{#1}}}
\newcommand{\tblref}[1]{\mbox{Table~\ref{#1}}}

\begin{document}

\title{Bit-Line Computing for CNN Accelerators Co-Design in Edge AI Inference}

\author{Marco~Rios, Flavio~Ponzina, Alexandre~Levisse, Giovanni~Ansaloni, David~Atienza

\IEEEcompsocitemizethanks{\IEEEcompsocthanksitem The authors are with the Embedded Systems Laboratory, École Polytechnique Fédérale de Lausanne (EPFL), Route Cantonale, 1015 Lausanne, Switzerland. \\E-mail: \{marco.rios, flavio.ponzina, alexandre.levisse, giovanni.ansaloni, david.atienza\}@epfl.ch.}

}
\IEEEtitleabstractindextext{%

\begin{abstract}
By supporting the access of multiple memory words at the same time, Bit-line Computing (BC)  architectures allow the parallel execution of bit-wise operations in-memory. At the array periphery, arithmetic operations are then derived with little additional overhead. Such a paradigm opens novel opportunities for Artificial Intelligence (AI) at the edge, thanks to the massive parallelism inherent in memory arrays and the extreme energy efficiency of computing in-situ, hence avoiding data transfers.
Previous works have shown that BC brings disruptive efficiency gains when targeting AI workloads, a key metric in the context of emerging edge AI scenarios. 
This manuscript builds on these findings by proposing an end-to-end framework that leverages BC-specific optimizations to enable high parallelism and aggressive compression of AI models. Our approach is supported by a novel hardware module performing real-time decoding, as well as new algorithms to enable BC-friendly model compression. Our hardware/software approach results in a 91\% energy savings (for a 1\% accuracy degradation constraint) regarding state-of-the-art BC computing approaches. 
\end{abstract}

}

\maketitle

\IEEEdisplaynontitleabstractindextext

%
\IEEEpeerreviewmaketitle

\IEEEraisesectionheading{\section{Introduction}\label{sec:introduction}}

\IEEEPARstart{T}{hanks} to their ability to extract abstract information from raw data acquisitions, Machine Learning (ML) algorithms such as Convolutional Neural Networks (CNNs) are fostering a revolution in multiple and diverse fields, ranging from personal mobility to health-care. Nevertheless, the increased accuracy of recent CNN models comes at the cost of massive memory requirements and intense workloads \cite{khan2020survey}.

These downsides are particularly important for edge devices running artificial intelligence algorithms, a scenario named edge AI in literature \cite{wang2020convergence}. Computational efficiency is key in edge AI because applications often have to abide to real-time constraints. Such constraints have to be met within tight computing and energy budgets, commonly orders-of-magnitude lower at the edge when compared to the cloud, thus requiring careful optimization of hardware and software.
The main avenues towards the optimization of ML workloads leverage the high level of parallelism and robustness of these algorithms. 

In the field of CNNs, parallelism is enabled by their structured and repetitive computing patterns, based on Multiply-ACcumulate (MAC) instructions, millions of which are employed to implement their convolutional  and fully connected  layers. Indeed, a high degree of data reuse is present both when convolving filters with activations (in convolutional layers) and when executing matrix-vector products (in fully connected ones). This characteristic can, as we do in this papers, be effectively harnessed by Single Instruction, Multiple Data (SIMD) computation strategies  to increase efficiency and performance \cite{chen2016eyeriss}.

Moreover, due to their robustness, CNNs can be optimized with very little, or no, accuracy drop, by either reducing the amount of required MAC operations or simplifying their computation. 
Quantization approaches advocate the use of fixed-point formats in contrast to floating-point, enabling the use of only few bits to represent the parameters (weights) and intermediate values (activations). Pruning strategies focus at a coarser granularity, seeking to skip weights and MAC computations that have little impact on the output quality. As detailed in Section \ref{sec:bg}, pruning and quantization are often combined in state-of-the-art edge AI strategies. 

Besides software optimization, the rise of edge AI has motivated the computer architecture and hardware research community to introduce dedicated designs. Approaches range from processors-based solutions, such as the ultra-low-power PULP multi-core in Rossi \textit{et al}. \cite{rossi2015pulp}, to custom accelerators, e.g., Eyeriss in Chen \textit{et al}.  \cite{chen2016eyeriss}. 
In this context, In-Memory Computing (IMC) architectures are particularly appealing, as  computation inside  memory  avoids  energy-expensive data movements in-between processing and storage components, while the parallelism made available by the regular structure of memory banks presents a good opportunity to support the SIMD patterns in CNNs.

A promising implementation of the IMC paradigm, that we focus on in this manuscript, is that of Bit-line Computing (BC) \cite{TC, rios2021running, rios19, ponzina2021flexible, darkside}. 
Based on conventional high-density SRAM, BC architectures can be seamless integrated in CMOS technology and target different levels in memory hierarchies. Moreover, BC supports a very high level of SIMD parallelism through the use of multiple subarrays at the same time. Finally, BC implementations require very little area overhead at the memory periphery to implement shift-add operations, which can then be chained to compute MACs. The above-mentioned approach is particularly beneficial when small-bitwidth operands are considered, as those require a reduced number of shift-adds. 



This work addresses the fundamental hardware/software co-design challenge by providing a holistic framework for the optimization, deployment, and execution of CNN models on a BC architecture for edge computing. 
We combine a novel BC-aware CNN optimization strategy with a highly optimized BC architecture. Both support fine-grained quantization and pruning in fully connected and convolutional layers. 
Moreover, leveraging the statistical distribution of weight values in CNNs, our methodology features a novel weight encoding strategy both during CNN optimization and in the BC hardware implementation. The strategy, named Generic Convolutional Weights (GCW) encoding,  uses few bits to encode weight values that appear more frequently, and a higher number of bits for those that are only rarely used, reducing model sizes by up to 4x in our experiments. 
A dedicated pipeline is in charge of decompressing the model representation at run time, without any impact on performance, converting it to a sequence of BC operations. Operations are then executed in parallel on multiple memory subarrays, greatly reducing run-time. 

In summary, the contributions of this paper are:

\begin{itemize}
    \item We present a synergic hardware and software framework that employs fine-grained bitwidths,  data compression, and in-memory parallel computing to support edge AI applications with a very high degree of energy efficiency.
    \item We detail the characteristics of a novel BC circuit implementation, able to effectively execute parallel in-memory MAC operations among operands of varying bitwidth. 
    \item We introduce a strategy to automatically map on BC subarrays the parameters of convolutional and fully connected layers, maximizing operations parallelism and data reuse.
    \item We present a CNN model compression, called GCW coding, to compress CNN models losslessly. Also, we describe a corresponding decoding circuit operating at run-time, with  little area and no timing overhead.

\end{itemize}
  
The remainder of the paper is organized as follows. 
Section \ref{sec:bg} discusses related works on CNN optimization/compression and on IMC architectures. 
Section \ref{sec:fw} introduces an overall view of the proposed framework. 
Section \ref{sec:sw_opt} focuses on the software optimization by detailing the CNN optimization methodology and the compression approach.
Next, Sections \ref{sec:bc} and \ref{sec:decompr} present the proposed BC memory array architecture and the pipeline circuit for GCW decoding, respectively. Section \ref{sec:cas} describes the mapping of convolutional and fully connected layers of CNNs on the BC architecture. 
Section \ref{sec:es} provides details on the experimental setup, while our achieved results are presented in Section \ref{sec:res}.
Finally, Section \ref{sec:con} concludes the paper. 

\section{Background}\label{sec:bg}

\subsection{Convolutional Neural Networks}
CNNs process input data employing a layer-based structure, where increasingly more abstract features are extracted in deeper layers. The compute-intense workload and large memory requirements of CNNs are mostly due to convolutional (CONV) and fully connected (FC) layers. For both layer types, the number of MAC operations is related to the size of input and output features, with several millions of MACs being common requirements in recent CNNs \cite{bianco18}. 

In CONV layers, three-dimensional matrices of CNN weights (named filters), are convolved over three-dimensional input feature maps, producing one output element for each filter position in the input. Conversely, FC layers (which are usually included after convolutional ones) compute linear transformations, multiplying the input feature vectors by the weight matrixes. FC and CONV layers have different data access patterns. In contrast to convolutions, weights in fully connected layers are used only once during an inference, because each column of the weight matrix multiplies the input vector to produce one output element. This characteristic is taken into account in the data mapping strategy discussed in Section \ref{subsec:IMCarch}. Recently, few other works have proposed methods to map large CNN layers in constrained IMC resources. As opposed to our focus on BC architectures, they consider crosspoint arrays \cite{song2017pipelayer, peng2019optimizing}, and provide solutions only for convolutional layers \cite{rhe2021vw}. 


\subsection{CNN models compression}

Pruning and quantization are the most common approaches exploiting the inherent redundancy in CNNs  to reduce their complexity,  hence supporting their deployment in constrained devices. In \emph{pruning}, either individual weights \cite{han2015learning} or entire convolutional filters \cite{he2020learning} are removed, achieving model compression higher than 10x \cite{molchanov2016pruning}. Instead, in
\emph{quantization} techniques, the weights and activations comprising CNNs models are represented using  low-bitwidths integer data representations instead of floating-point numbers \cite{reagen2018ares}. Quantized CNN models have a smaller memory footprint than floating-point ones. Furthermore, they use less complex integer hardware for computing MAC operations, which also reduces energy requirements.
It has been shown that 8-bits quantization can be usually implemented without affecting the initial CNN accuracy \cite{denkinger2019impact}.


A further approach to compress CNN models is that of weight \emph{encoding}. It is often applied after quantization, as low-bitwidth representations constrain the set of admissible values. Encoding can be implemented employing different strategies. Codebook-based strategies limit the number of unique weights and store them in small look-up tables (i.e., named codebooks), encoding the baseline model into a set of binary indexes that address specific code-words\cite{jain2020symmetric}. 
Another weight encoding approach is to leverage the statistical distribution of weight values to compress their representation, employing shorter code-words for the most used values and longer code-words for rarer ones\cite{han2015deep},\cite{encode2}.

Since encoding weight values does not change CNN models, but only operates on data representation, it does not degrade accuracy. On the other side of the coin, run-time decoding may introduce overheads  in time, area, and energy requirements. The authors of \cite{han2015deep} use Huffman codes to index a codebook storing a constrained set of CNN weight values. While our GCW strategy (detailed in Section \ref{sec:sw_opt}) has some similarities with respect to Huffman coding, it does not require explicit look-up tables, minimizing the cost of its hardware implementation.

\subsection{IMC architectures}
\label{subsec:IMCarch}

In-Memory Computing (IMC) relies on smart memories to perform computing tasks in a highly parallel way. IMC is especially attractive for applications that are computationally and memory-intensive, while presenting regular patterns and few (or no) data-dependent branches or loop-carried dependencies. CNNs do present these characteristics, and are therefore well-suited for IMC execution.

IMC architectures can be broadly divided into two categories. Crosspoint architectures employ meshes of variable resistances to encode weights and arrays of analog-to-digital converters to read-back results \cite{ielmini2020device}. They can efficiently perform matrix-vector multiplications, but are prone to noise due to manufacturing variability, temperature, and voltage fluctuations \cite{ ielmini2018memory , alevisse2017}. Moreover, they require the challenging co-integration of digital, analog, and non-volatile technologies.

A second strategy, that we focus on in this paper, is that of Bit-line Computing (BC) \cite{tcam}.
This approach can be readily integrated into existing SRAM memories and hierarchies by adding the capability to concurrently activate multiple memory words and employing sense amplifiers to perform bit-wise operations. 
As the name suggests, bit-line Computing relies on the behavior of the bit-lines (BL and $\overline{\text{BL}}$) when two Word-Lines  (WLs) are activated in the same cycle clock. 

\begin{figure}
    \centering
    \vspace{-0.3cm}
    \includegraphics[width = 0.8\linewidth]{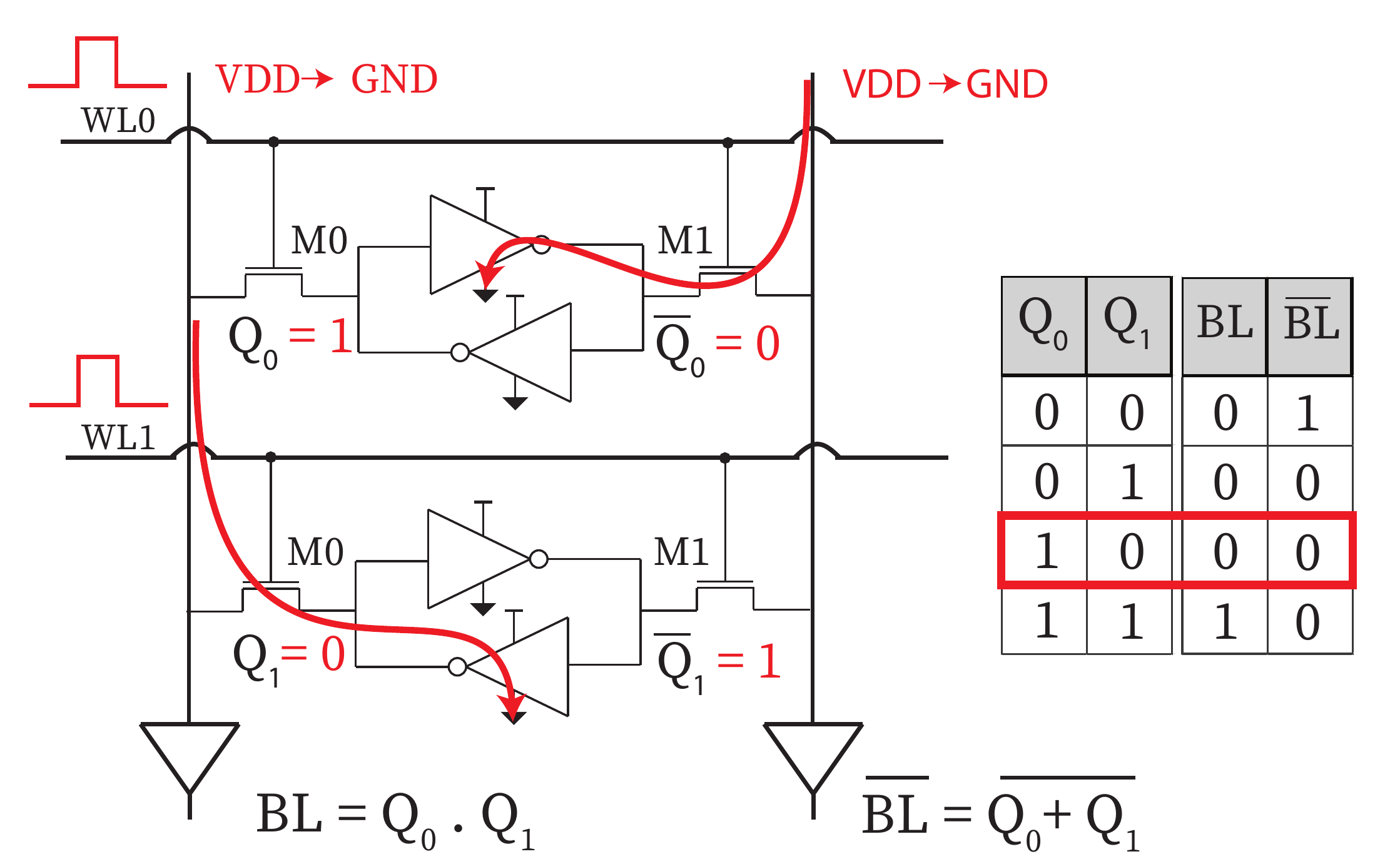}
    \caption{Bit-line computing concept. Two word-lines are activated in the same clock cycle. The discharge of the bit-lines result in the bit-wise operations AND and NOR between the accessed words.}
    \vspace{-0.4cm}
    \label{fig:bc}
\end{figure}

Such a behavior is exemplified in \figref{fig:bc}, which shows two SRAM bit-cells (each realized as two cross-coupled inverters) connected to the same BLs. When the two WLs of the cells are activated, if either one or both of the cells store the value '0', the voltage on the BL wire discharges to the ground through one or both of the cells access transistor (M0) and the NMOS of their inverter. Notice that the only case in which BL remains at Vdd is when both cells store the value '1'. Thus, the BL connection behaves as the logic AND gate. Dually, the negated BL signal ($\overline{\text{BL}}$) only retains a high voltage if both cells store the value '0' (hence $\overline{\text{Q}_0}$ and $\overline{\text{Q}_1}$ are both '1'), implementing the functionality of a NOR gate.

This setup can induce data-corruption due to undesirable currents flowing between the bit-cells. In order to address this issue, \cite{tcam} reduces the operating frequency and \cite{drc2} adopts 10T bit-cells. However, these solutions either decrease the performance or imply large area overheads. Recent works \cite{TC, rios2021running} have shown that such operations can be reliably performed at high clock frequencies if the activated WLs belong to different subdivisions of SRAM arrays, termed Local Groups (LGs). Only memory cells in the same LG share short-distance vertical connections (local bit-lines, LBLs). During a read access, cell values are propagated via LBLs to the LG Periphery (LGP) circuit  which features sense amplifiers and read/write ports. In-memory bit-line operations are then performed after those sense amplifiers, effectively protecting memory cells from data corruption.

Further logic is then employed at the array periphery to derive additions from bit-wise operations. 
Finally, multiplication instructions are performed as a sequence of shift-adds between two memory words storing a) one of the two multiplication operands and b) the accumulated partial results (details are provided in Section \ref{sec:bc4}). 

This strategy differentiates between the two inputs of a multiplication, as only one of them (\emph{In-Memory Operand}, IMO) resides in the memory array, while the other (\emph{Broadcasted Operand}, BO) is streamed in the memory one bit at a time. Indeed, a single BO can be streamed to multiple memory subarrays at once to perform multiplications in parallel sharing one operand (hence its name). As shown in Table \ref{tab:operands}, due to the differences in access patterns and to maximize data reuse, in CONV layers we consider weights as BOs and activations as IMOs, while for FC layers we operate the opposite choice.

Although shift-adds require several cycles to execute multiply or MAC instructions, high performance can be achieved in BC arrays   by (i) word-level parallelism inside a single subarray, (ii) partitioning the SRAM into subarrays to perform parallel operations, (iii) workload optimization to reduce the bitwidth of streamed BO operands, reducing the cycle-count of multiplications. 
The effectiveness of (i) and (iii) is highly influenced by quantization and pruning, as they  determine the number of shift-adds required by a multiplication. Nonetheless, current works do not explore this interdependence in detail. As an example, \cite{reagen2018ares} adopts a fixed 16-bits representation for activations and an 8-bits one for weights. We instead show that important efficiency gains can be harnessed when employing more aggressive stances such as heterogeneous per-layer schemes, and that these can be supported with little hardware overhead.

\begin{table}
\centering
\caption{Assignment of operands for fully connected and convolutional layers}
\begin{tabular}{l|c|c}
\diagbox{Operand}{Layer type}
&\makebox{\textbf{CONV}}&\makebox{\textbf{FC}}\\\hline
\textbf{Multiplicand} & \multirow{2}{6em}{Activations} & \multirow{2}{6em}{Weights}\\
In-Memory Operand (IMO) & &\\ \hline
\textbf{Multiplier} & \multirow{2}{6em}{Weights} & \multirow{2}{6em}{Activations} \\
Broadcasted Operand (BO) & & \\
\end{tabular}
\vspace{-0.3cm}
\label{tab:operands}
\end{table}


\section{Framework Overview}\label{sec:fw}

\begin{figure}
    \centering
        \vspace{-0.3cm}
    \includegraphics[width = 0.85\linewidth]{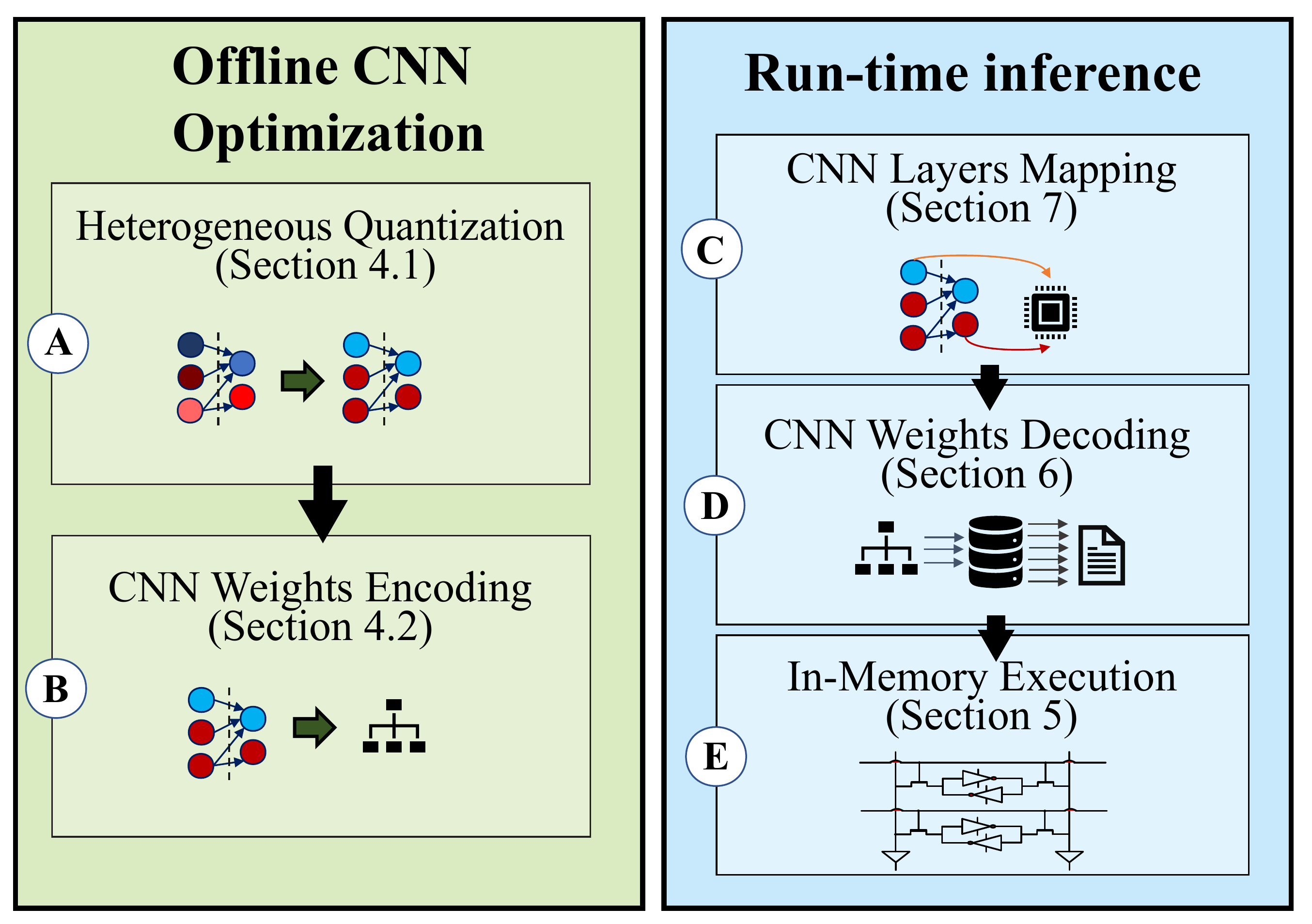}
    \caption{Overall view of our HW-SW co-design framework, showing algorithmic optimizations (left), mapping, and execution on BC hardware (right).}
        \vspace{-0.3cm}
    \label{fig:fw}
\end{figure}

\figref{fig:fw} provides a bird's eye view of our framework. Starting from a (floating-point, non-optimized) description of a CNN the framework provides a pathway to co-optimize the CNN implementation and the BC computing array executing it. It also optimizes the mapping of software onto hardware.

Application-level optimizations, detailed in \secref{sec:sw_opt}, combine non-uniform quantization schemes and encoding methods to obtain an optimized model that can be efficiently executed in-memory.
In more detail, we first include a resource-aware CNN quantization strategy that reduces workload and memory requirements (\figref{fig:fw}.A). This stage consists in an iterative approach, where the bitwidth of weights and activations in convolutional and fully connected layers is optimized while abiding by an accuracy threshold.
Then, due to the characteristic distribution of CNN weights,  GCW encoding further compresses the quantized CNNs (\figref{fig:fw}.B). GCW uses smaller bitwidths for the weight values appearing more frequently and larger bitwidths for those appearing only sporadically.

Because of the typically large size of intermediate features in  CNNs, we decompose convolutions and the matrix-vector operations in FC layers into smaller computing blocks (\figref{fig:fw}.C). This tilling process, focus of \secref{sec:cas}, allows large CNN models to be accelerated in limited-sized memories, minimizing the number of data transfers while  exploiting parallelism to increase performance.

When processing convolutional layers, weights are decoded at run-time during execution, with minimal overhead (\figref{fig:fw}.D). We describe the circuit performing weights decoding in \secref{sec:decompr}.  The decoder first translates parameters to their two's complement representations, and then converts them to a sequence of BC instructions, which govern the execution of our memory arrays (\figref{fig:fw}.E). The design of the BC arrays and their features, including support for heterogeneous quantization, are detailed in \secref{sec:bc}. 

\section{Algorithmic-level CNN optimization} \label{sec:sw_opt}

Algorithmic optimizations (\figref{fig:fw}.A) aim at decreasing the workload and memory requirements of a  CNN model with BC-aware transformations. The optimization process consists of two stages: first, a heterogeneous quantization and pruning step reduce the bitwidth and the number of weights and activations in convolutional and fully connected layers. Next, the quantized weights are encoded using variable-bitwidth codes, further reducing memory requirements. 

\subsection{Heterogeneous Quantization}\label{sec:mixed_precision}

Per-layer quantization enables aggressive model compressions. Indeed, layers in CNN may (and, usually, do) have different degrees of robustness, with quantization-sensitive layers requiring larger bitwidth for activations and parameters. Heterogeneous schemes have therefore the potential to reach better trade-offs between accuracy and model size. However, they also expose a much larger design space than uniform alternatives.

\begin{figure}
    \centering
    \includegraphics[width = 0.85\linewidth]{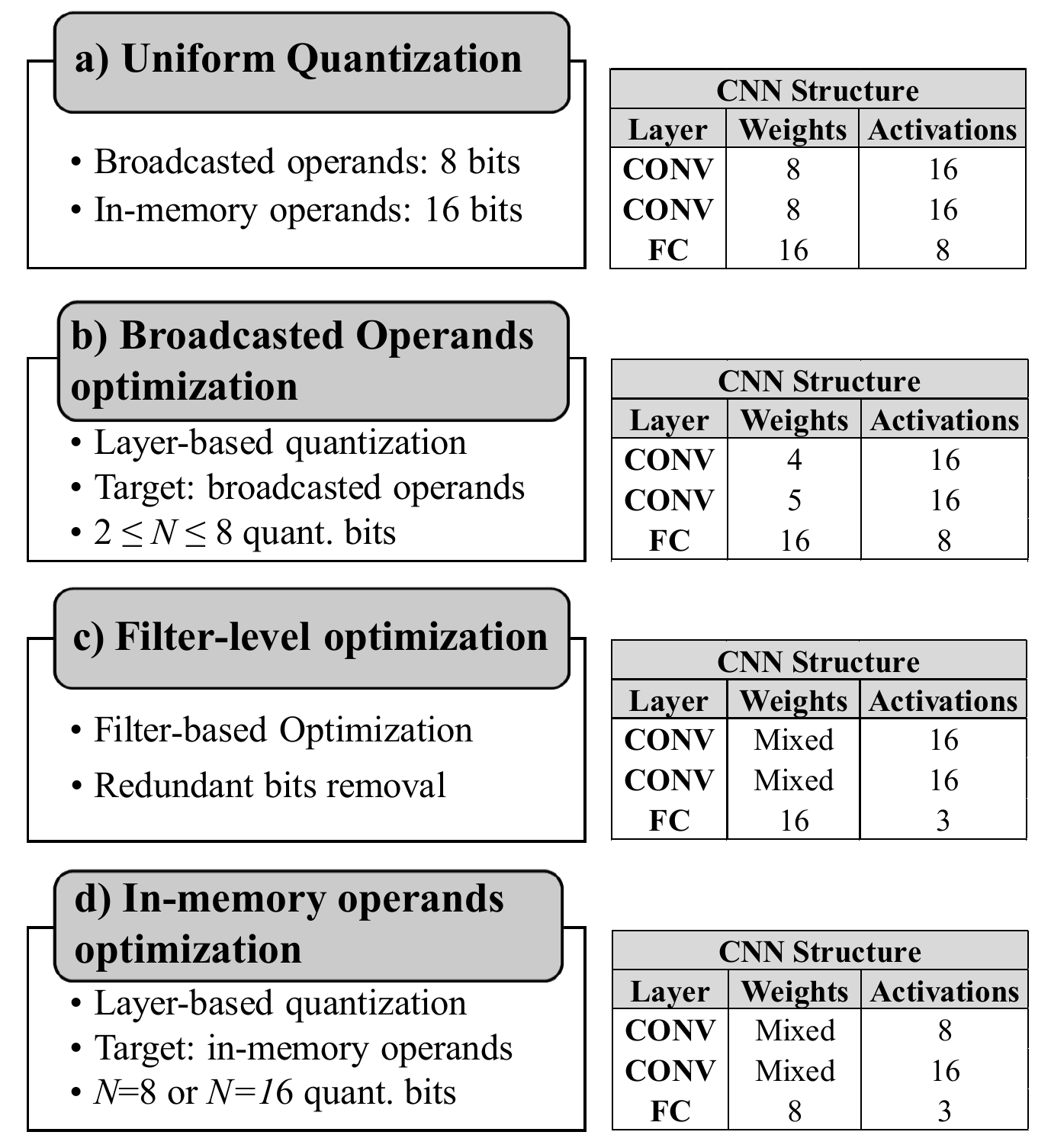}
    \vspace{-0.3cm}
    \caption{Workload-aware quantization and pruning methodology (left). Running example (right).}
    \vspace{-0.5cm}
    \label{fig:flow}
\end{figure}

To navigate it, we introduce an iterative process that reduces the size of in-memory and broadcasted operands (IMOs and BOs, as defined in Table \ref{tab:operands}), according to the scheme in \figref{fig:flow}. 
As a running example, the figure considers in its rightmost part a running example referring to a CNN with two convolutional and one fully connected layers, which we will follow in the rest of this section.

The input models in our optimization flow are homogeneously quantized CNNs, employing 16-bits IMOs and 8-bits BOs (\figref{fig:flow}(a)). As shown in \cite{reagen2018ares}, CNNs at these quantization levels have indistinguishable accuracies with respect to floating-point implementations.
In the first optimization step (\figref{fig:flow}(b)), the bitwidth of the BOs is independently tailored for each layer.
To this end, we attempt to reduce bitwidths starting from the layer having the highest number of MAC operations (and, therefore, the highest potential for savings). 
Then, we retrain the network for a small number of epochs and check the obtained accuracy of the new configuration. 
If the accuracy degradation exceeds a user-defined threshold, we backtrack (we considered 1\% and 5\% maximum degradations for the experiments in Section \ref{sec:res}) .
In a similar fashion, we then iteratively target the layers having the second, third, etc. most numerous MAC operations. 
Once all layers have been processed once, we further try to reduce the bitwidth of the BOs in the highest-workload layer from which we haven't previously backtracked.
The iteration continues until no further bitwidth reductions are possible in the BOs of any layer.

This phase is followed by a filter-level optimization, which focuses on convolutional layers only. We observe that a large amount of CNN filters do not use the entire value range when representing weights, especially after the above-mentioned aggressive quantization. 
As illustrated in \figref{fig:flow}(c), we hence drop, without loss of accuracy, the most significant bits (MSbs) on a per-filter base if allowed by weight ranges, correspondingly scaling convolution outputs. 
For example, if the value range of the BOs in a filter is $\subset [-0.25,0.25)$, 2 MSbs can be dropped, and outputs should be divided by $2^{Dropped\_bits} = 4$. 
Additionally, in this stage, filters having all weights equal to zero are entirely deleted.

The last step of the optimization flow (\figref{fig:flow}(d)) performs the tailoring of the IMOs bitwidths. It leverages the word-level parallelism supported by the BC arrays (as described in Section \ref{sec:bc5}). 
Similarly to the approach followed for BOs, we attempt to reduce the bitwidth of IMOs on a per-layer basis. 
However, quantization steps are coarser in this case, as they must abide to the sub-word formats supported in hardware. In our experiments, we admit 1x16-bit and 2x8bit sub-words, with the latter resulting in 2x reduction in execution time\footnote{While in principle our approach could be extended to 4 bits or 2 bits per word, such settings would incur in unacceptably large accuracy degradations.}.

\subsection{Generic Convolutional Weights Encoding} \label{sec:hu}

\begin{figure}
    \centering
    \includegraphics[width = 0.9\linewidth]{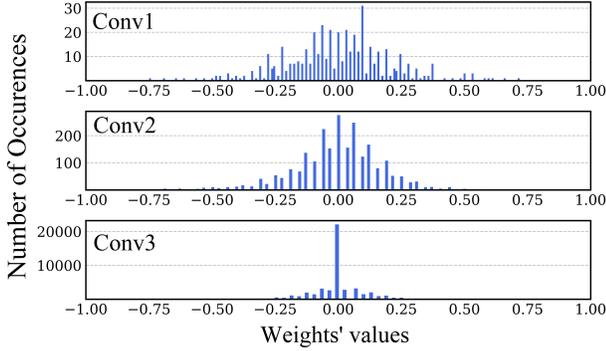}
    \vspace{-0.2cm}
    \caption{Weights distribution in the three convolutional layers of LeNet-5.}
    \vspace{-0.3cm}
    \label{fig:weights_distribution}
\end{figure}

\begin{figure}
    \centering
    \vspace{-0.1cm}
    \includegraphics[width = 0.65\linewidth]{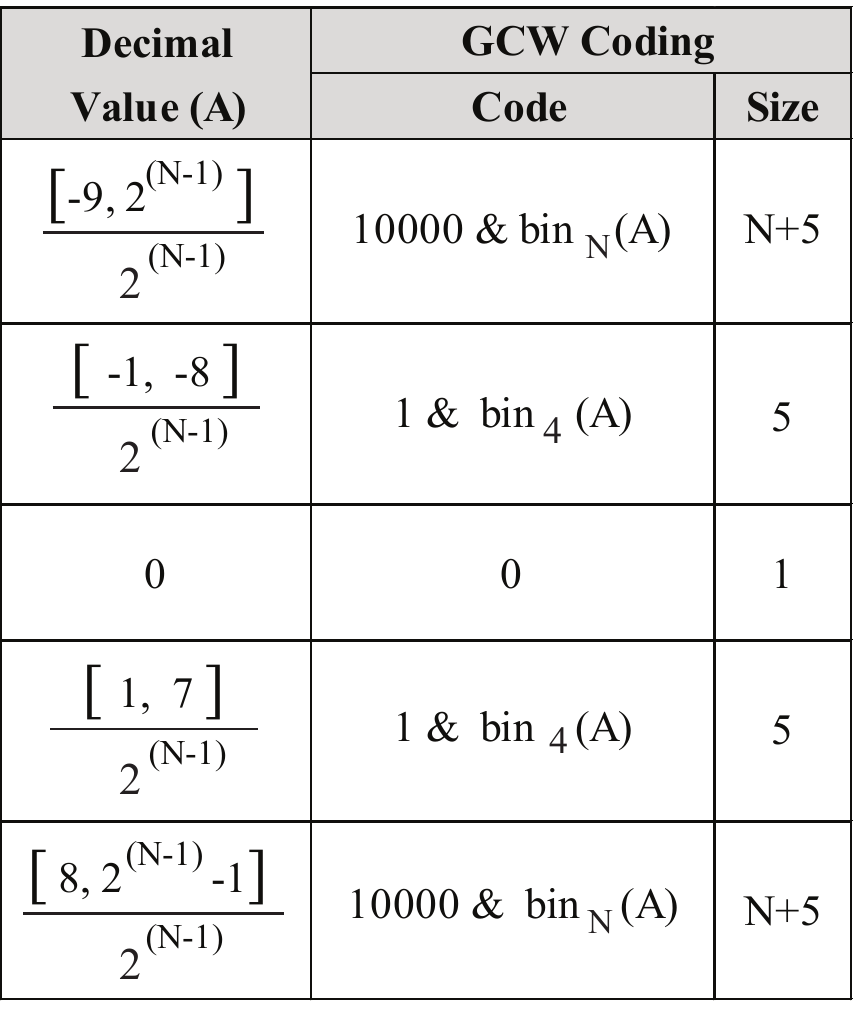}
    \caption{Generic Convolutional Weights coding scheme, where $N$ indicates the quantization level values before coding. The code uses fewer bits to represent the most frequent symbols. }
    \vspace{-0.3cm}
    \label{fig:GCW}
\end{figure}

GCW compression further reduces the memory required to represent parameters in convolutional layers. 
Our strategy leads to an efficient implementation of run-time decoding, which we describe in \secref{sec:decompr}. It  takes advantage of the limited set of values that can be assumed by quantized weights, as well as their characteristic statistical distribution.

We observe that weights in quantized CONV layers predominantly assume the value "0", even after
removing  filters only containing zero values. This trend is illustrated in \figref{fig:weights_distribution}, which provides as examples the weights distribution in the three convolutional layers of LeNet-5 \footnote{We obtained similar results for the other benchmark CNNs in \secref{sec:res}.}.

Additionally, narrow distributions centered around zero indicate that small (quantized) weight values appear much more frequently than larger-magnitude ones. 
These findings open the path for an encoding scheme that, similarly to Huffman coding, employs code-words of variable length. Highly occurring values are coded using low-bitwidth representations, while less frequent values are mapped to higher bitwidth codes. Given the distributions of CNN weight values, this choice translates into employing the minimal bitwidth (1 bit) for the value "0", small bitwidths for the values close to zero and large bitwidths for values of high magnitude. 



The GCW encoding scheme is illustrated in \figref{fig:GCW}, with $N$ being the number of quantized bits. 
Weights are assumed to be normalized in the range   $[-1, 1)$. They are divided into five intervals, symmetric with respect to zero (as shown in the leftmost column). 
Values close to zero (other than zero itself) are represented with 5-bit code-words. A 1-bit prefix is appended to the two's complement 4-bit representation of the corresponding value.

Other, seldomly used,  code-words are derived by appending a fixed 5-bit prefix to their $N$-bit representation.

For $N=8$, 13 bits are required to represent large-magnitude values, but only 5 bits for low-magnitude ones (and only 1 bit for the value '0'). Note that, for $N<5$, long code-words are never generated and the coding only generates different code-words lengths for zero and non-zero weights.

\section{Bit-line Computing Architecture} \label{sec:bc}

\subsection{BC array organization} \label{subsec:bc_org}

\begin{figure}
    \centering
    \includegraphics[width = 0.96\linewidth]{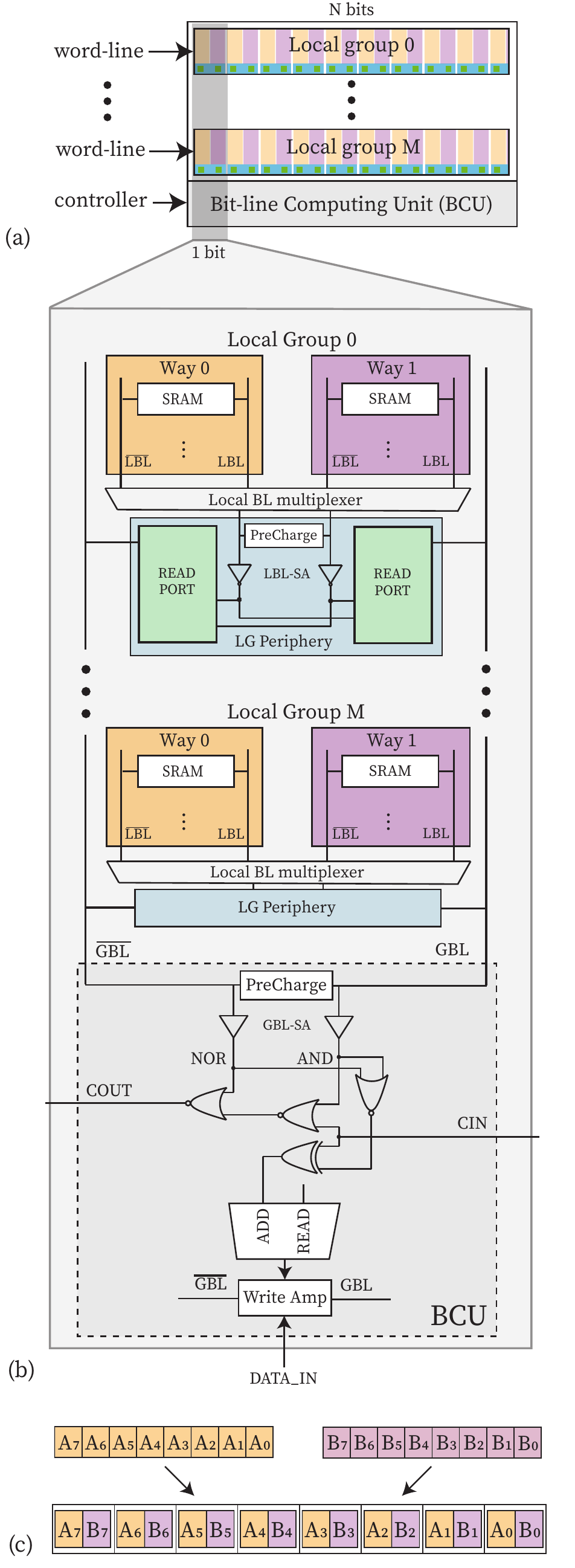}
    \caption{(a) Bit-line Computing subarray. (b) Highlight of one bit-column depicting the Local Group circuitry and Bit-line Computing Unit. (c) 8-bits two-ways interleaving.}
    \label{fig:subarray}
\end{figure}

As commonly done in standard SRAMs, the BC array architecture is divided into several subarrays. Each subarray is capable of performing a MAC operation between a broadcasted operand (which must be the same for all subarrays) and an in-memory operand, local to each subarray. 
\figref{fig:subarray}(a) shows the subarray organization. The Word-Lines (WLs) are evenly separated into Local Groups (LGs), which contain the memory cells and the LG Periphery (LGP). The role of the LGP is to electrically isolate SRAM cells from each other, allowing in-memory operations among words in different local groups at high speed without data corruption.  LGPs logic also performs bit-wise negation and shift operations, as detailed in the following. At the bottom of the subarray, the Bit-line Computing Unit (BCU) ripples a carry signal among bit columns, allowing the implementation of additions. Subarray operations are governed by a small controller, global to the entire array, which decodes BC operations and properly activate WLs. 

\figref{fig:subarray}(b) highlights a single bit-column of the subarray, in which bit-cells are organized in Local Groups (LGs). Moreover, words in a LG are arranged with way interleaving (i.e., the same bit of multiple words are placed next to each other)\footnote{For clarity, in \figref{fig:subarray}(b) we show an example with only two ways}. Bit-cells belonging to the same local group and the same way share, via access transistors, share the same Local Bit Lines (LBLs). LBLs in a LG are connected to a multiplexer that acts as a way selector, allowing to connect LBLs to the sense amplifiers of the LG periphery and, through its read ports, to global bit-lines (GBLs). Finally, GBLs interface with the BCU, which, besides standard SRAM read and write ports, implements a 1-bit adder, rippling the carry in and out from/to neighboring bit-columns.

Our design based on way interleaving allows the memory array to be designed with high-density push-rule 6T bit-cells, while still maintaining enough clearance for the design of the LGP and BCU blocks, which must be vertically aligned with storage cells.

\subsection{In-Memory Operations}  \label{sec:bc3}

The design illustrated above supports both standard read-write and in-memory operations, as detailed next.

\subsubsection{Write} 
Write operations are performed by setting up the data on the GBLs through the write amplifier, which chooses data from two sources: the external input (DATA\_IN in \figref{fig:subarray}) or the result of a performed BC operation. Thereon, the LGP write port \footnote{Not depicted in \figref{fig:subarray}} transfers the data from the GBLs to the LBLs. Finally, the WL is activated, switching on the access transistor and allowing the data to  be written into the target SRAM bit-cell. 

\subsubsection{Read} 
Both LBLs and GBLs are pre-charged to \texttt{Vdd} before reading a word from the SRAM memory. Then, the WL is activated, switching on the access transistor and allowing one of the LBLs to discharge to the ground. Conversely, the other holds its charge, depending on the value stored in the bit-cell. The LBL Sense Amplifier (LBL-SA) consequently asserts a logic value and this signal is connected to the read port. GBLs assume the voltage value of the LBLs connected to the read ports, and the GBL Sense Amplifier (GBL-SA) asserts such value and outputs the read word.

\subsubsection{BC Instructions} 

\begin{figure}
    \centering
    \includegraphics[width = \linewidth]{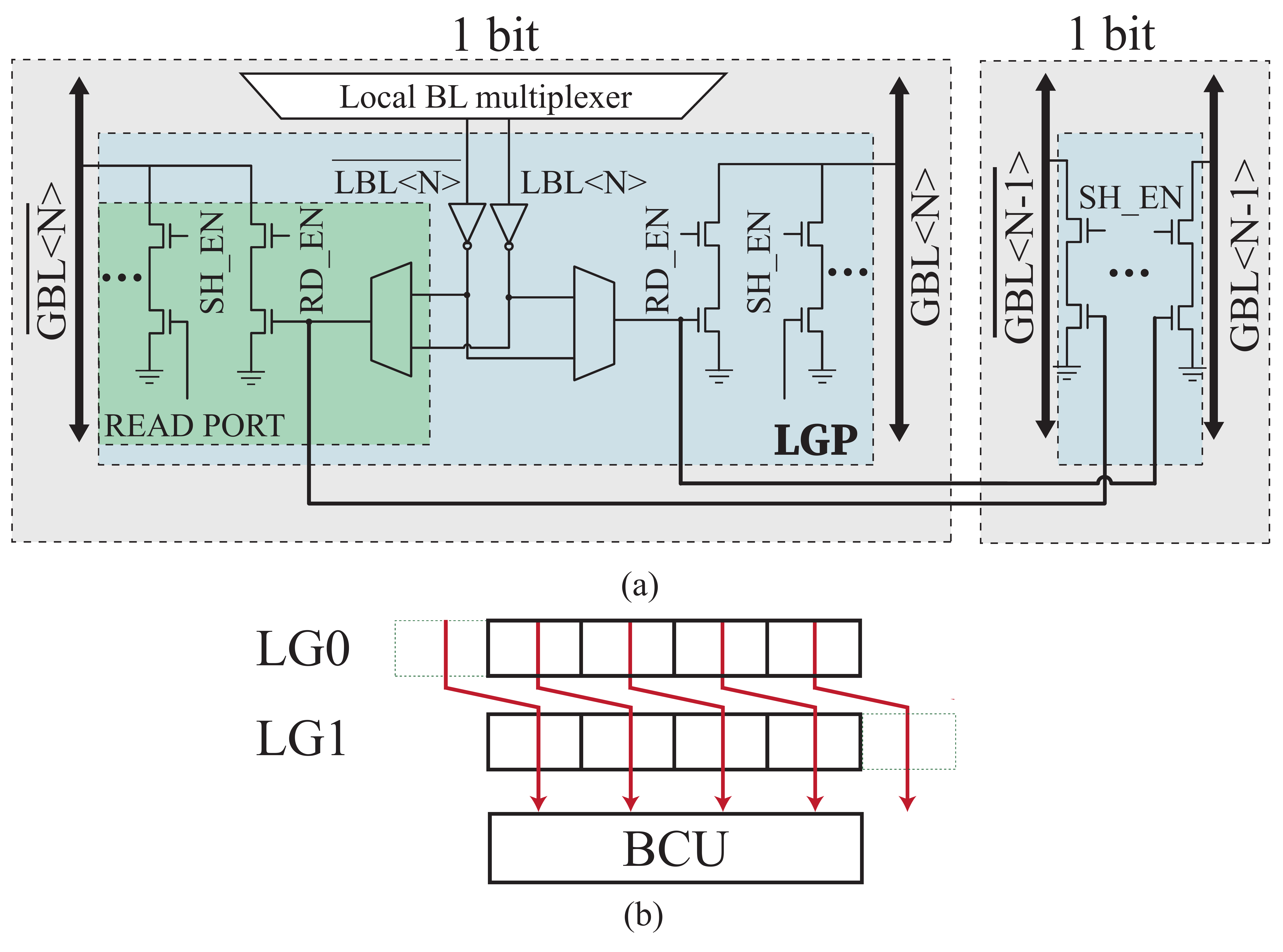}
    \vspace{-0.4cm}
    \caption{(a) Local Group Periphery circuit with extended read port, enabling shift and negation. (b) Data path of embedded shift. }
    \label{fig:readport}
    \vspace{-0.4cm}
\end{figure}


BC instructions  always assume the format of \texttt{add(OP1, OP2)}, with the operands being allowed to be individually shifted and/or negated. OP2 can be set to zero if only shift/negation of one operand is required. The operation outcome can be either output from the memory or written back, which requires an additional cycle. \texttt{shift} and \texttt{negation} operations are performed on each operand in the LGP by the read ports, allowing the operands to be individually shifted/negated before the addition is performed.

\figref{fig:readport}(a) presents a schematic of the read port circuitry. It  offers two paths to discharge the GBLs to the ground, where each path is composed of two NMOS in series. Conventional read operation uses the path controlled by the Read Enable (RD\_EN) signal and the LBL-SA output of the same bit. Shifts instead rely on the path controlled by the Shift Enable (SH\_EN) signal and the LBL-SA output of a \emph{neighbour} bit. 
This operation is defined as Embedded Shift (ES), since it is performed inside the LGP.
Indeed, single-cycle, multiple-shifts operations can be supported in read ports by adding additional discharge paths from LBL-SA of farther bit columns, accomplishing design with a varying Number of Embedded Shifts (NES).

The LGP also embeds 2-to-1 multiplexers connected to both outputs of the LBL-SA. This design allows bit-wise negations to be supported in the LGP, as for this operation the multiplexers invert the LBL and $\overline{\text{LBL}}$. At the subarray level, values can consequently be arithmetically negated (i.e., in two's complement representation), by performing bit-wise negation in the LGP and asserting the carry-in of the least significant bit in the BCU.

\subsection{Multiplications}  \label{sec:bc4}

\subsubsection{Multiplication among binary numbers with partial products}

In general terms, multiplications of two operands can be decomposed into partial products, which are summed up to retrieve the full product.  Each partial product is found by multiplying each digit of the multiplier with the multiplicand and shifting the result to the left based on the position of this digit. In binary systems, the digits of the multiplier are either zero or one, thus, the partial products are either zero or the left-shifted multiplicand. 

\begin{figure}
    \centering
    \includegraphics[width = \linewidth]{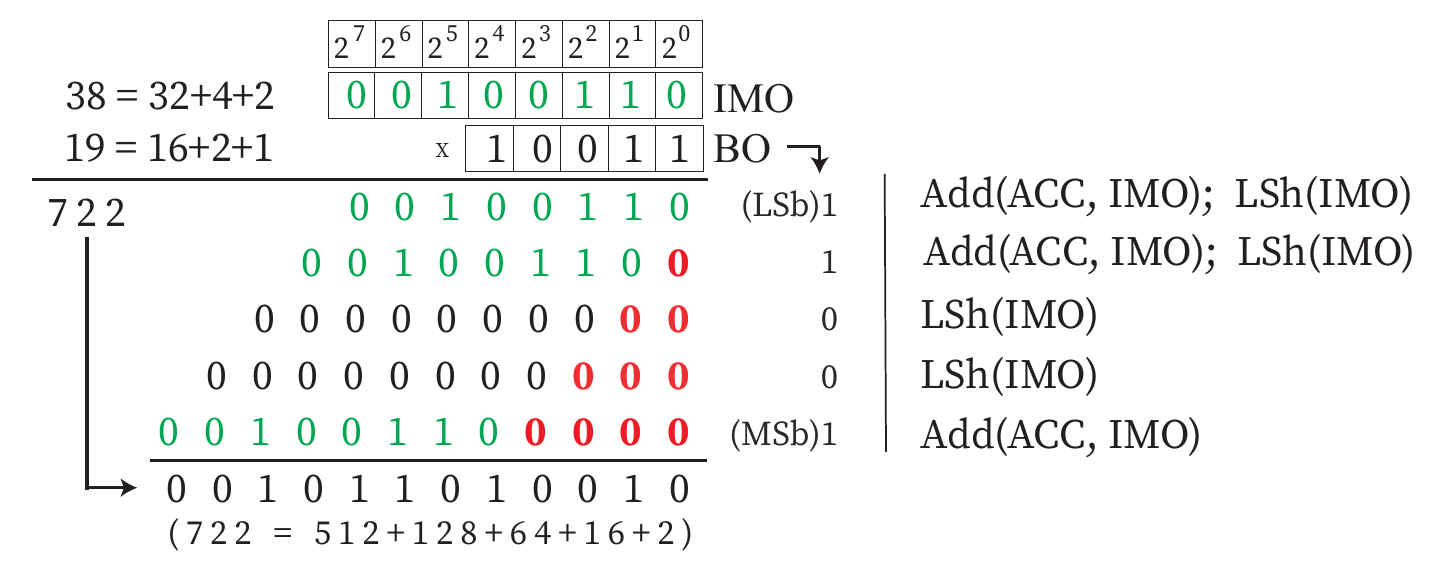}
    \caption{Binary integer multiplication example, between the IMO \mbox{(00100110$_2$ = 38$_{10}$)} and the BO \mbox{(10011$_2$ = 19$_{10}$)}, the multiplication is decomposed in shift-add instructions. LSh operations are left-shifts.}
    \vspace{-0.4cm}
    \label{fig:bim}
\end{figure}

\figref{fig:bim} illustrates an example that shows a binary multiplication of two input operands, the IMO \mbox{(00100110$_2$ = 38$_{10}$)} and the BO \mbox{(10011$_2$ = 19$_{10}$)}.  \figref{fig:bim} also shows the required operations, where ACC is the partial products at each iteration. It can be noticed that operations are only performed with the IMO and product vector ACC, while each bit of the BO dictates which operation should be done at each step.

\subsubsection{Multiplication in the BC array}

The approach to multiplication illustrated above can not be executed as-is in the BC architecture. First, in \figref{fig:bim}, the product \mbox{(1011010010$_2$ = 722$_{10}$)} requires more bits (10) compared to both of the \mbox{operands (5 and 8-bits)}, which would result in an overflow in our BC scenario. Second, no support is provided for representing negative numbers, common in AI applications. 

These two issues are addressed by supporting MAC operations in fixed-point (instead of integer) format, among signed numbers encoded in two's complement. We represent values using 1 bit for the integer part, and $n$  bits for the fractional part (a format commonly indicated as $Q1.n$). Thus, the values are always in the range $[-1, 1)$. Since two's complement is used, the most significant bit is `1' for  negative numbers and `0' for positive ones.

Multiplication on the BC architecture can then be performed by employing sequences composed by the following two instructions:
\begin{enumerate}[label=(\roman*)]
    \item Addition of two operands, both arithmetically right-shifted (i.e., with sign extension) by one bit. One operand is the partial product ACC, while the second is either 0 or the IMO, depending on the BO bit. 
    \item Addition between two operands, where one operand is the partial product ACC and the other is either 0 or the 2's complement of the IMO.
\end{enumerate}

Instruction (i) is performed at every iteration step from bit 0 to bit $N-1$ of the BO, while operation (ii) is used for bit $N$ of BO (its most significant bit).

\begin{figure}
    \centering
    \includegraphics[width = 0.9\linewidth]{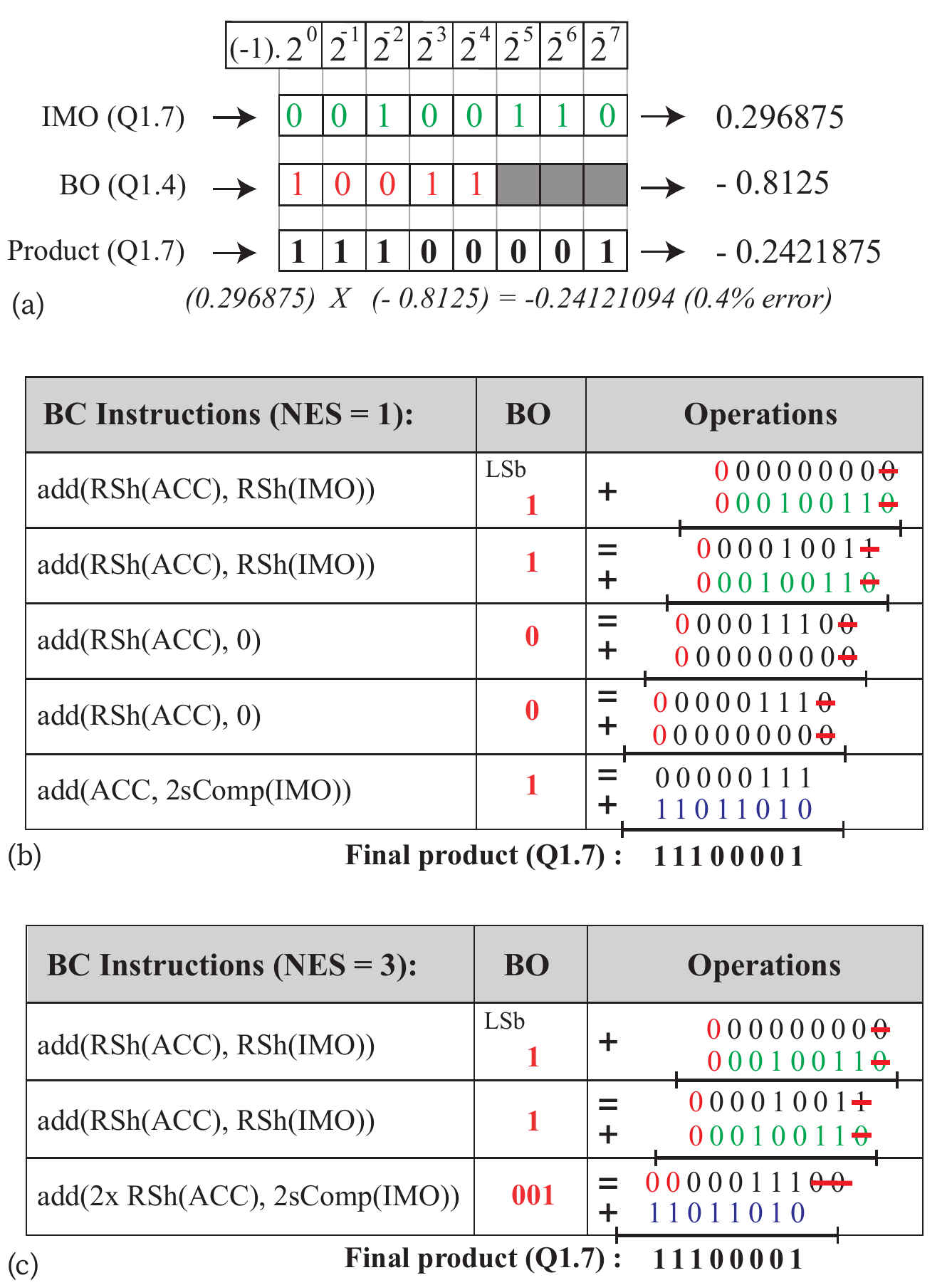}
    \caption{(a) Two's complement fixed-point multiplication example between the IMO expressed in Q1.7 and the BO expressed in Q1.4. The BC instructions for (b) NES = 1 and (c) NES = 3. RSh operations are right-shifts.}
    \label{fig:twoscomplementmultiplication}
    \vspace{-0.5cm}
\end{figure}

\figref{fig:twoscomplementmultiplication}(a) shows an example of a multiplication among two two's complement numbers in the Q1.7 and Q1.4 formats, with the result in Q1.7. Binary digits are the same as the ones depicted in \figref{fig:bim}, however, in  \figref{fig:twoscomplementmultiplication} the associated values are now  \mbox{IMO = 00100110$_2$ = 0.296875$_{10}$} and \mbox{BO = 10011$_2$ = -0.8125$_{10}$}. 

\figref{fig:twoscomplementmultiplication}(b) shows the operations required to perform the multiplication, in the case of only supporting shifts of 1 bit (NES = 1). Notice that the bitwidth of the partial products does not increase at each iteration, and no overflow occurs. Instead, truncations induce small errors in the computed product (0.4\% in the example). 

Crucially, all the listed iterations include operations which are supported in the BC array, and each iteration can be executed in a single clock cycle. 
As presented in \figref{fig:readport}, right-shifts are provided by read ports in Local Groups, and can execute concurrently with additions, since the BCU computes additions at the subarray periphery. Similarly, the addition of the partial product with the two's complement of the multiplicand, also requires one clock cycle as the bit-wise negation is done in the LG periphery, while the addition and the assertion of the carry-in is executed in the BCU.

\subsubsection{Using Multiple Embedded Shifts}

The support for  multiple embedded shifts (NES$>$1) can effectively speed-up computation, at the cost of added complexity in the read port. Indeed, in the case of \mbox{NES = 2}, 2 bits of the BO can be processed in one iteration, provided that the first one is equal to zero (``00'' and ``01''), since in this case only one addition, or none, is required after two shifts. For \mbox{NES = 3}, BO sequences of bits with two leading zeroes (``000'' and ``001'') can be processed in a single clock cycle.
\figref{fig:twoscomplementmultiplication}(c) shows the BC instructions to execute the same multiplication as \figref{fig:twoscomplementmultiplication}(b), but on a BC array with \mbox{NES = 3}. In this configuration, only 3 operations are required (instead of 5 when \mbox{NES = 1}), resulting in a speed-up of 1.67x. A large number of supported NES can provide diminishing benefits, as long sequences of zeroes in the BO can be rare. Overly large NES values may also be detrimental, as they impact the complexity of LGP read ports \cite{rios19}.

\subsection{In-Memory Parallelism} \label{sec:bc5}

\begin{figure}
    \centering
    \includegraphics[width = 0.85\linewidth]{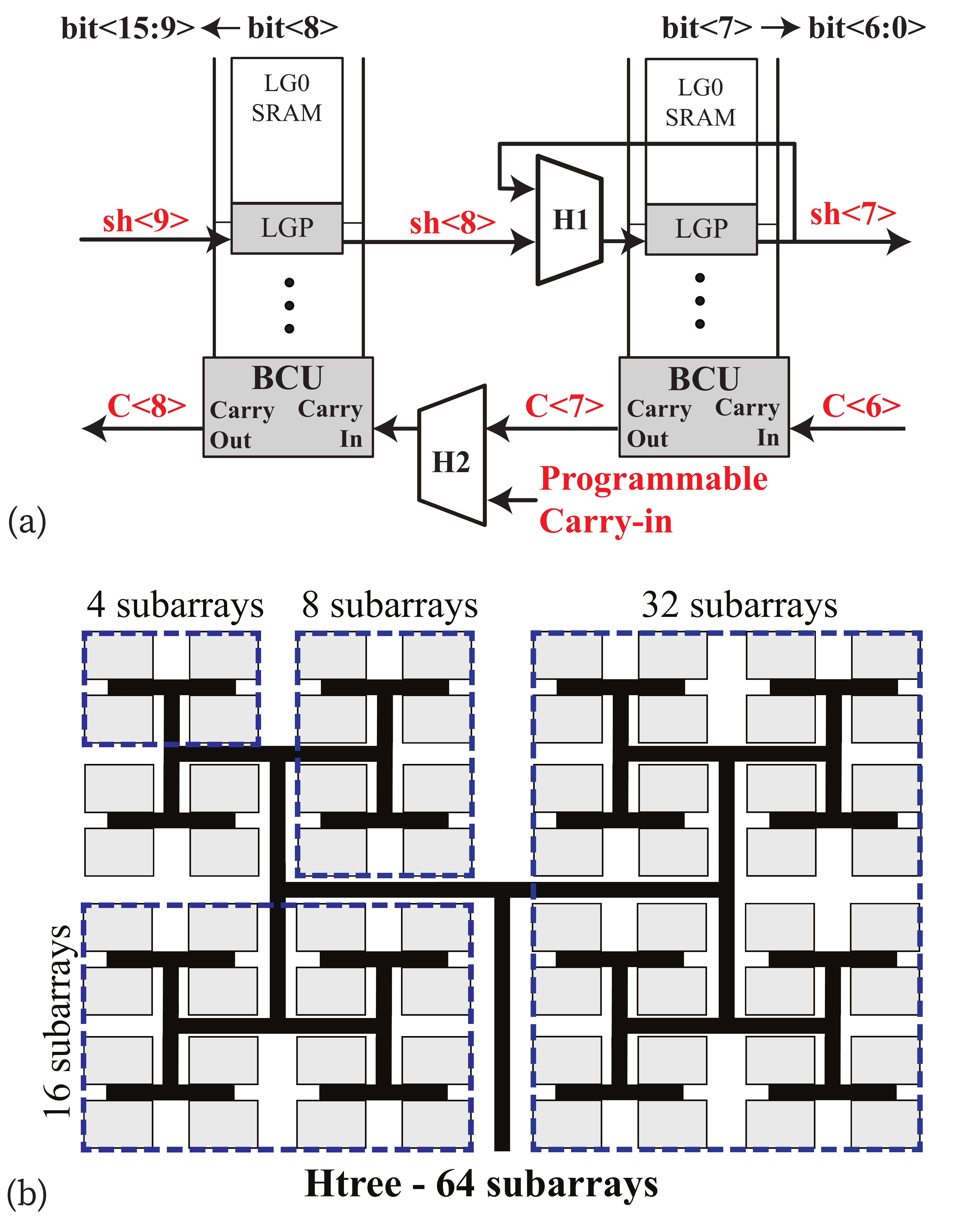}
    \caption{(a) Word-level parallelism support for 1x16bit and 2x8bit modes, considering NES = 1. (b) H-tree organization.}
    \label{fig:subarray2}
    \vspace{-0.5cm}
\end{figure}

\subsubsection{Word-level Parallelism}

The BC architecture allows to store in each memory location either a single 16-bit word (in Q1.15) or two 8-bit ones (each in Q1.7)
\footnote{Word-level parallelism does not impact normal (non-BC) reads and writes.}.
In-memory instructions can hence be executed on 1x16bit or 2x8bit word formats, enabling in the latter case two multiplications simultaneously in a single subarray. 

To implement word-level parallelism, connections between the $7^{th}$ and $8^{th}$ bit-column are configurable. In particular,
right shift operations on the $7^{th}$ bit-column can either be connected to the $8^{th}$ bit-column (in 1x16bit mode), or to itself (to implement sign extension in 2x8bit mode). Similarly, the carry-in of the $8^{th}$ bit-column can be either connected to the $7^{th}$ bit-column (1x16bit mode), or dictated by the performed operation to implement two's complement (2x8bit mode). 

Hence, two additional multiplexers are required (H1 and H2 in \figref{fig:subarray2}). The first one,  at LG peripheries, connects or disconnects the shift-right signal from the $8^{th}$ to the $7^{th}$ bit-column, while the second, in BCUs, operates similarly for the carry signal between the $7^{th}$ and the $8^{th}$ bit columns.



\begin{figure*}
    \centering
    \vspace{-0.3cm}
    \includegraphics[width = 0.8\linewidth]{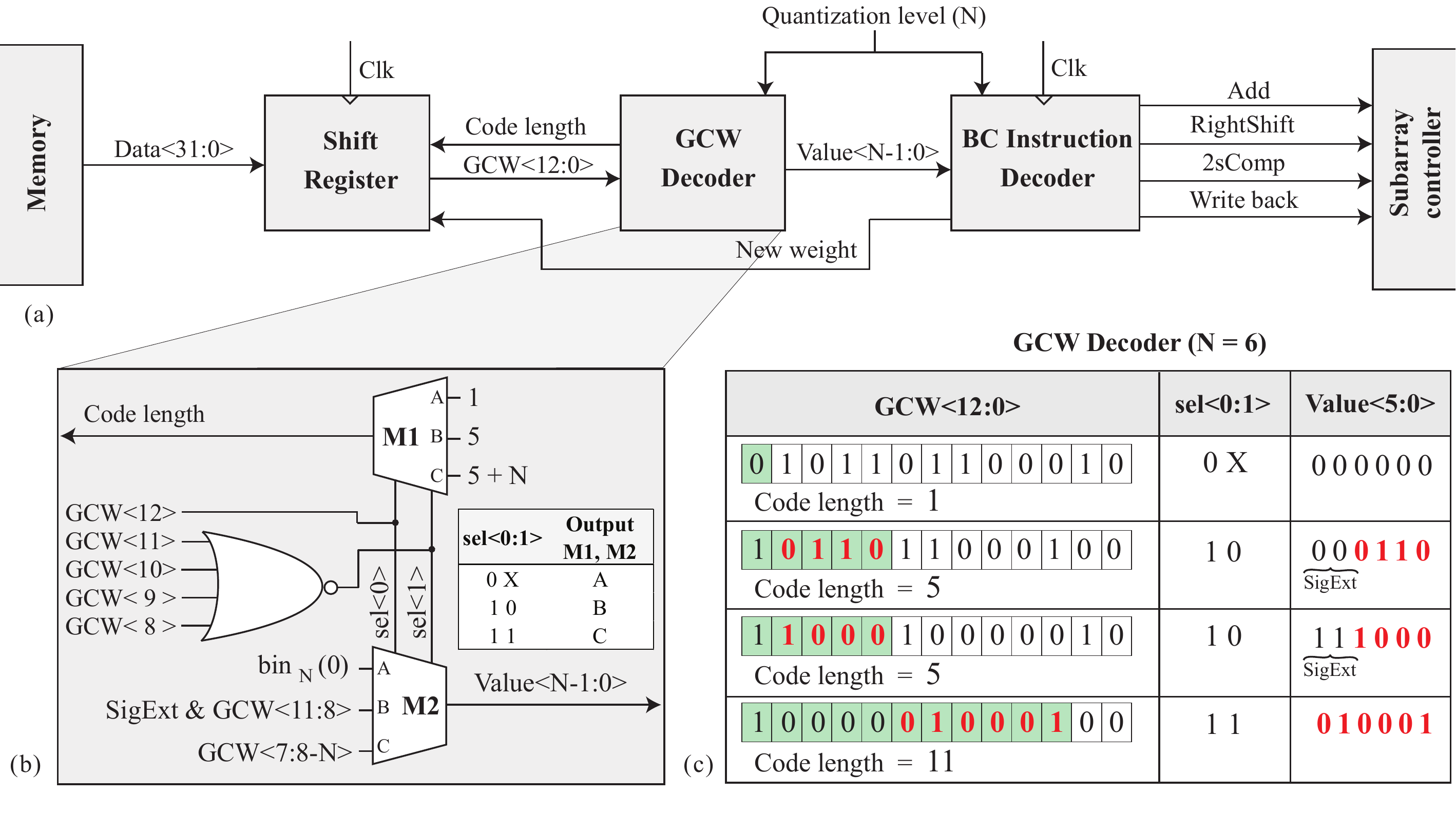}
    \caption{(a) Block diagram of the pipeline circuit to extract BC operations from the compressed CNN model. (b) GCW decoder circuit-level design (c) GCW decoding example with N = 6.} 
        \vspace{-0.1cm}
    \label{fig:hufdec}
        \vspace{-0.3cm}
\end{figure*}

\subsubsection{Array-level Parallelism}

Multiple subarrays are connected in an H-tree configuration, as shown in Figure \ref{fig:subarray2}(b). This organization ensures that all the signals transmitted from/to the array periphery have the same distance to all the subarrays, equalizing the delays and minimizing critical paths. 

The H-tree interconnect is active before and after computation to transfer IMO inputs to subarrays, and retrieve results. 
During computation, the H-tree broadcasts one BC instruction related to BO bits to all subarrays at each clock cycle, exploiting the reuse of BOs in multiple MAC operations to achieve a high degree of parallelism. 
Importantly, this approach allows a fine-grained flexibility in the bitwidth of BOs, which may assume arbitrary values, leading to a fine-grained control of trade-offs between model cost (size, energy, time-per-inference) and accuracy.

\section{Real-time Data Decompression}\label{sec:decompr}

To reduce the size of CNN models, hence memory requirements, convolutional weights are stored in an encoded form, according to the Generic Convolutional Weights (GCW) representation introduced in Section \ref{sec:hu}. At run-time, a pipeline circuit decodes them, deriving the corresponding BC instructions. \figref{fig:hufdec}(a) depicts the pipeline stages of the decoder. First, a shift register reads a memory word containing multiple GCW-encoded weights, possibly of different bitwidths. Then, the GCW decoder decompresses weights into their Q1.$n$ representation. Finally, the BC instruction decoder converts these values into a set of BC instructions, which are broadcasted to the subarrays.

\subsection{Shift Register} 
For each convolutional filter, GCW-encoded weights form a bit-stream where each value is represented with 1 bit (for the "0" value), 5 bits (for values close to 0), or up to 13 bits (otherwise). Hence, the shift register, which is filled by 32-bits memory words at a time, usually holds multiple GCW code-words.
When the decoder requires a new weight, it examines the first 13 bits of the shift register  (\texttt{GCW$<$12:0$>$} in \figref{fig:hufdec}(a)), determining which bit-field blue(first bit, first 5 bits, or first 5+$N$ bits, where $N$ is the quantization level) contains the next code-word and advances the shift register according to the code length. 
The shift register is re-filled when less than 13 bits remain in its buffer, concatenating  memory words. Code-words can hence cross the boundary of two subsequent memory words, preventing memory under-utilization.

\subsection{GCW Decoder}
The GCW decoder decodes the weights values according to their quantization level $N$, as illustrated in \figref{fig:GCW}. 
It analyzes \texttt{GCW$<$12:0$>$}, searching for specific bit sequences. 

The circuit-level design of the GCW is depicted in \figref{fig:hufdec}(b). It comprises two multiplexers 3-to-1 (M1 and M2), which share the same selection signal (\texttt{sel$<$0:1$>$}). The signal \texttt{sel$<$0$>$} is directly connected to \texttt{GCW$<$12$>$}, while \texttt{sel$<$1$>$} is the output of a 4-input NOR gate (\texttt{GCW$<$11:8$>$}). When \texttt{sel$<$0$>$} = 0, the weight value is zero, encoded using a 1-bit code. Instead, when\texttt{sel$<$0$>$} = 1, \texttt{sel$<$1$>$} controls the output of the multiplexers. If the \texttt{GCW$<$11:8$>$} bits are different from ``0000'', they represent a small-magnitude weight encoded using a 5-bit code. The corresponding value is GCW$<$11:8$>$, sign-extended to $N$ bits (\texttt{SigExt}). Finally, if \texttt{GCW$<$11:8$>$} = ``0000'', the code-word represents a large-magnitude weight, whose value is encoded in the \texttt{GCW$<$7:8-N$>$} bits using a variable code length that varies from 9 to 13.

\figref{fig:hufdec}(c) illustrates an example of run-time decoding  considering $N=6$. In the first one, \texttt{GCW$<$12$>$} = 0. Consequently, the weight value is zero, encoded using 1 bit. Thus, the shift register shifts one position, allowing the next weight to be decoded.
Next, in the second example, \texttt{GCW$<$12:8$>$} = "10110". The GCW decoder extracts the last 4 bits and concatenates them with ``00'' to form a 6-bits binary value. The third example shows a similar case, but, as the read value is negative, ``11'' is concatenated. The last example depicts the case where the GCW decoder finds the sequence ``10000''. The binary value is then retrieved in \texttt{GCW$<$7:2$>$}. 

\subsection{BC instruction decoder} 
The BC instruction decoder converts the weights' bits into BC instructions, defining the \texttt{RightShift}, the \texttt{Add}, and the \texttt{2sComp} signals. It also governs the write back signal to store the result of in-memory operations in the BC arrays. The sequence of shift-add operations implementing a MAC is entirely skipped when a "0" weight value is decoded, which results in energy and performance gains, as discussed in \secref{sec:res}. 


\section{Mapping CNNs to BC Arrays} \label{sec:cas}
The presented BC architecture effectively accelerates the execution of convolutional and fully connected layers, whose workload dominates the execution of CNN models (e.g., they constitute more than 98\% of the run-time in the benchmarks in Section 8).
When deploying a CNN layer onto a BC array of given dimensions, primary goals are the reduction of data transfers between the subarrays and the periphery, and the maximization of parallelism when computing MACs. 
To this end, we developed an automated operations scheduler, which distributes workloads to the BC architecture considering hardware constraints, such as the number of available subarrays and their size, as well as application characteristics, including the type of CNN layer (convolutional or fully connected) and its geometry. 
Because large CNN layers may not entirely fit the limited memory size of the available BC arrays, the operations scheduler divides each layer into smaller blocks, or tiles, processed in parallel by each subarray.
We detail next the specific mapping strategy for either CONV and FC layers.

\subsection{Convolutional Layers}
\begin{figure}
    \centering
    \includegraphics[width = 0.9\linewidth]{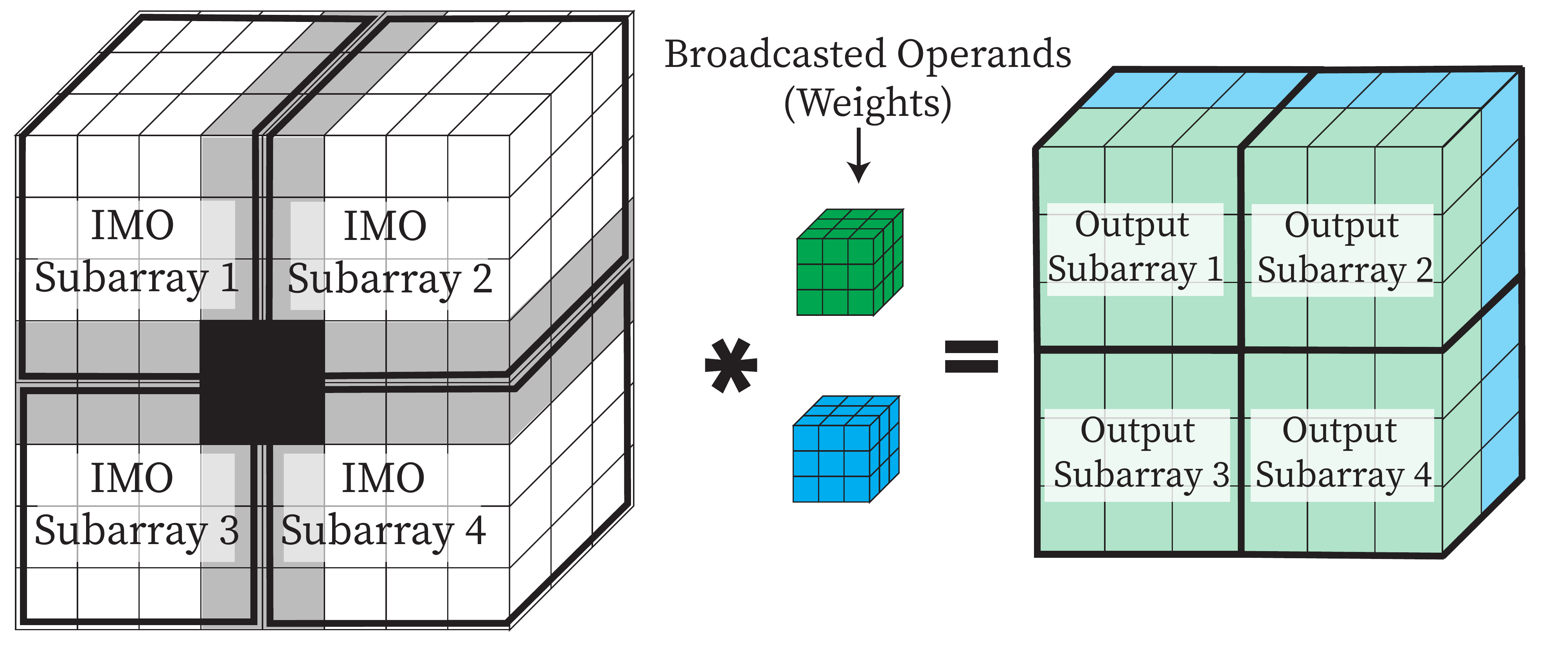}
    \caption{Convolutional layer activation tilling and transferred to different subarrays, while weights are converted into BC operations. }
    \label{fig:conv1}
\end{figure}

\begin{figure}
    \centering
    \includegraphics[width = 0.9\linewidth]{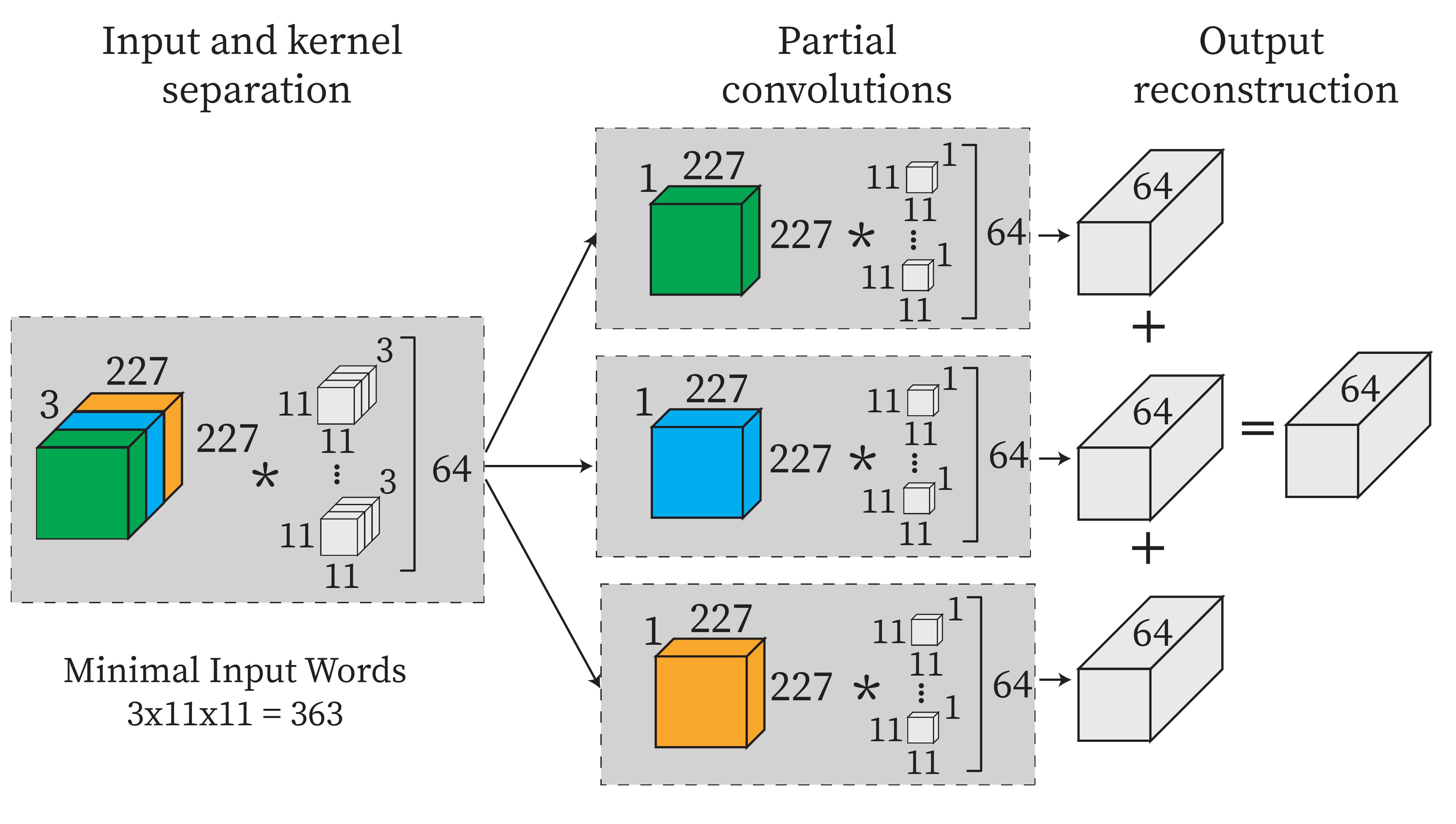}
    \caption{Example of convolution layer inputs that surpass the capacity of a subarray, this layer is deployed with partial convolutions, which requires an extra in-memory operation to reconstruct the output activation. }
    \label{fig:conv2}
\end{figure}

As presented in \tblref{tab:operands}, for convolutional layers  activations are  in-memory operands, while weights are broadcasted to subarrays in the form of BC instructions. \figref{fig:conv1} shows an example of how the layers are split into tiles and assigned to subarrays. In this example, input activations consist in a matrix of 8x8x3 that is convoluted by two filters of 3x3x3. Each output element is generated by computing the Hadamard product, or element-wise product, between a filter and a fraction of the input matrix having the same size, and summing up all the elements of the output Hadamard product. 
Sliding the filter over the entire input matrix allows the evaluation of different output elements. Hence, the convolution of the input features with one filter results in a bi-dimensional output (of size 6x6x1 in the example). Then, multiple filters compute different convolutions, producing a three-dimensional output (6x6x2 \figref{fig:conv1}). 

To parallelize this computation pattern, a filter is used to convolve multiple input memory regions at the same time, storing the data pertaining to each region as the in-memory operands of different subarrays. Note that this parallelization strategy involves broadcasting the filter weights to subarrays. A subdivision in four memory regions is depicted in the example in \figref{fig:conv1}, each computing a quarter of the output (a 3x3x2 matrix). 
Due to border effects, memory regions must partially overlap. In the example, the elements in grey are transferred to two subarrays, while the elements in black are transferred to all of them. For practical subarray sizes, nonetheless, this effect is marginal, even if the size of tiles reduces when the number of employed subarrays increases. 

For large CONV layer sizes and small subarray size, even small tiles require more input words to compute a single output value than the subarray storage capacity. For example, considering the first layer of AlexNet that has 64 filters of size 11x11x3, requiring 363 input words to calculate a 1x1 output for all the 64 filters (1x1x64), which surpasses the BC implementation detailed in \secref{sec:es} (320 16-bits words per subarray). In these cases, partial convolutions are performed, as shown in   \figref{fig:conv2}. Consequently, filters are decomposed in the depth direction. Then, partial convolution results are retrieved with BC operations as before. Finally, the outputs of the partial convolutions are merged with additional in-memory operations.

\subsection{Fully Connected Layers}

Fully connected layers compute $Y$ outputs from $X$ inputs by performing a vector-matrix multiplication with a $X\times{}Y$ weight matrix $W$. In contrast to CONV layers, each weight is only used once at each inference in FC ones. Therefore, it cannot be used as a broadcasted operand. Instead, inputs do exhibit data reuse, as all $X$ components are employed in the computation of each output. 
Therefore, we store weights as IMOs in memory subarrays for FC layers, and employ inputs as BOs. 

\begin{figure}
    \centering
        \vspace{-0.3cm}
    \includegraphics[width = 0.9\linewidth]{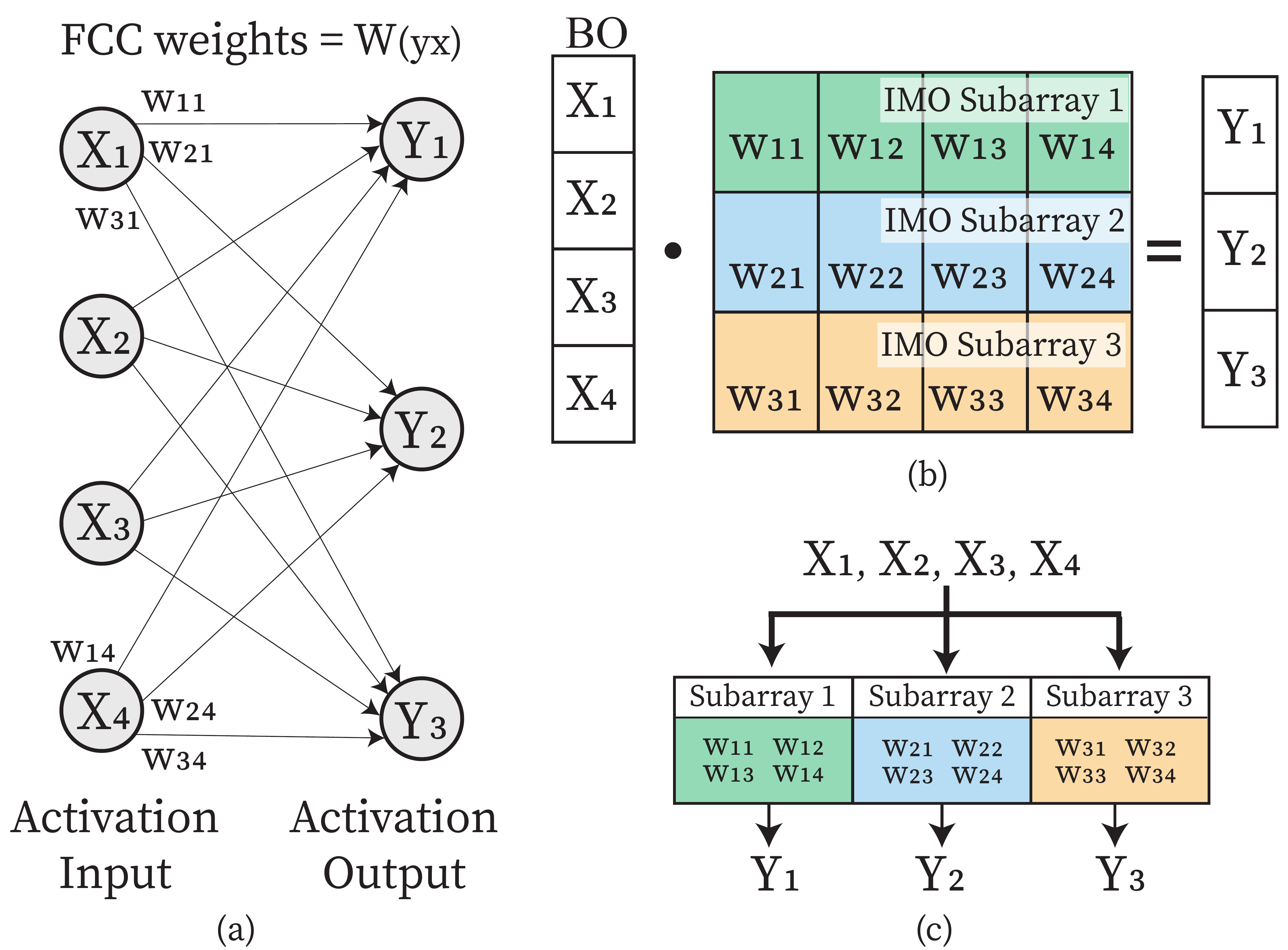}
    \caption{Example of FC layer represented in (a) graphical form and (b) matrix multiplication. (c) Data mapping to execute this workload on a BC array. }
    \label{fig:fcc}
        \vspace{-0.3cm}
\end{figure}

This mapping is exemplified in \figref{fig:fcc}(a), where the input vector $X$ has four elements and output $Y$ has three elements. Consequently, twelve weights are required for the layer in the example. \figref{fig:fcc}(b-c) depicts how this layer is mapped into three subarrays to calculate the three outputs (i.e., one output per subarray). The MAC operations are performed between the common input $X$, which is broadcasted to the subarrays as BC instructions, and the stored weights $W$. Thus, all the three outputs are calculated in parallel.  

Such FC mapping is not amenable to data encoding strategies such as GCW, because broadcasted operands are activation values computed at run-time during inference. Nonetheless, FC layers exhibit usually smaller workloads than CONV ones and are not, in general (as in the benchmarks considered in \secref{sec:res}), run-time bottlenecks.

\section{Experimental Setup} \label{sec:es}
\subsection{Evaluation on CNN benchmarks}

We evaluate our architecture and optimization framework on several edgeAI benchmarks of different complexity to demonstrate the effectiveness of our approach in a significant range of applications: we consider LeNet-5~\cite{lenet} on \mbox{CIFAR-10} ~\cite{krizhevsky2009learning} and AlexNet~\cite{alexnet}, VGG16~\cite{simonyan2014very}, MobileNet~\cite{mobilenet} and Xception~\cite{chollet2017xception} on the CIFAR-100 dataset~\cite{krizhevsky2009learning}. 
Accuracy values for various CNNs and optimization levels are retrieved using PyTorch~\cite{paszke2019pytorch} and the quantization functions described in~\cite{jacob2018quantization}. 

CNNs are first trained using floating-point precision for 200 epochs, obtaining accuracies in line with the state-of-the-art. 
Similar to \cite{ponzina2021flexible}, models are then homogeneously quantized to 16-bits in-memory operands and 8-bits broadcasted operands, and refined for 20 additional training epochs. This configuration, which has no accuracy loss with respect to employing floating-point weights and activations, is assumed as the starting point for further optimizations using the proposed methodology.  

To establish a baseline, we iteratively repeat quantization and retraining to homogeneously reduce, down to 2 bits, the bitwidth of the operands streamed into the subarrays. Then, we retrain the models for five  fine-tuning epochs at each step. 
Five epochs are also run when applying our heterogeneous approach, as described in Section \ref{sec:sw_opt}, after each BOs and IMOs optimization steps (phases (b) and (d) in \figref{fig:flow}).

\subsection{Circuit-level characterization}

\begin{figure}
    \centering
    \vspace{-0.3cm} 
    \includegraphics[width = 0.8\linewidth]{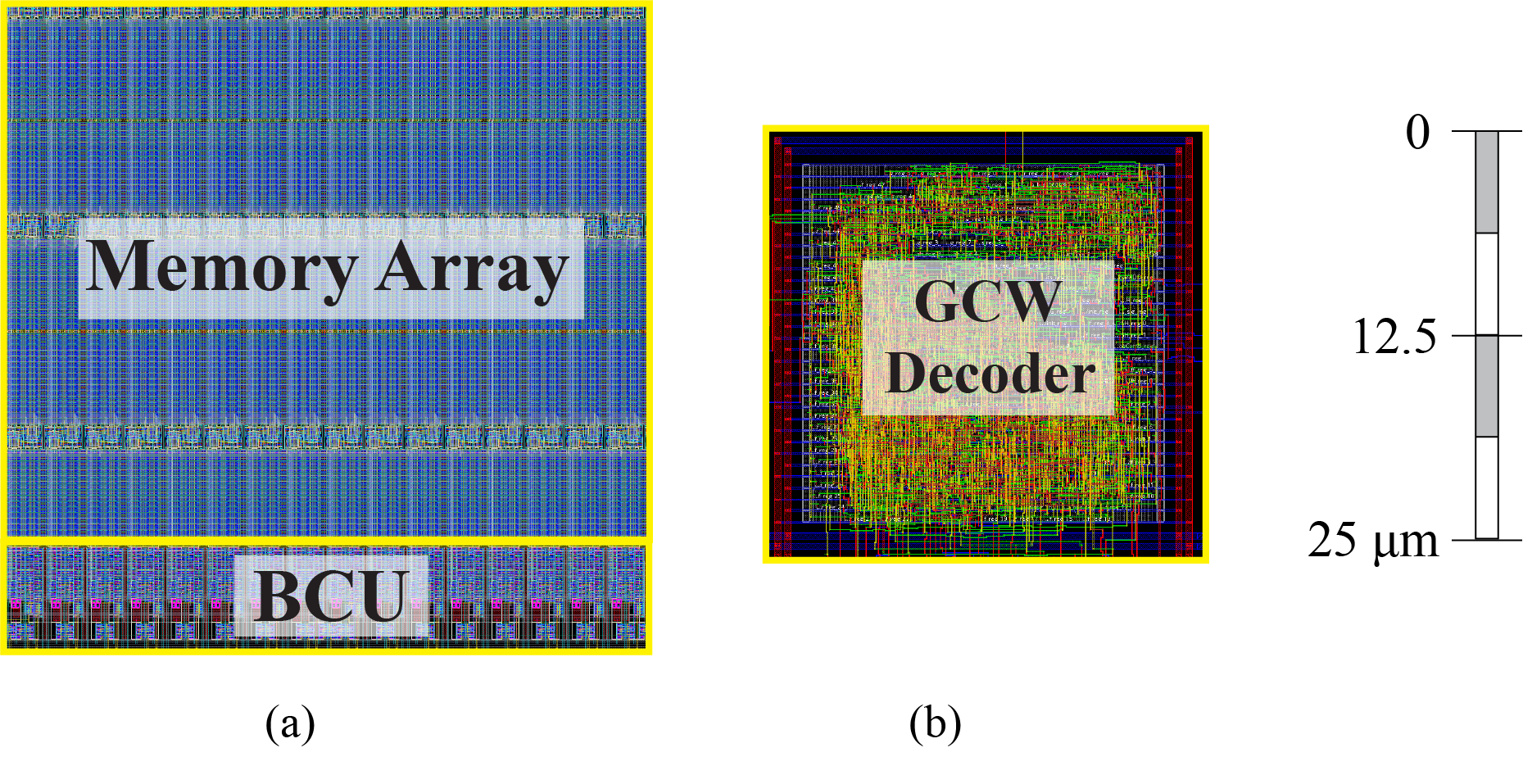}
    \vspace{-0.2cm} 
    \caption{Layout of (a) semi-custom design of the decoder circuit and, (b) full-custom design of one memory subarray.}
    
    \label{fig:layout}
\end{figure}

As a test vehicle for our experiments, we consider a BC architecture composed of a varying number of subarrays. Each subarray contains 5 LGs of 32 rows each, totalling 160 memory rows. Each row is composed of 2 interleaved 16-bits words. Therefore, the subarray stores 5120 bits (640 bytes). From a layout perspective (\figref{fig:layout}(a)), the memory cell array is organized into 160 rows and 32 columns. It can store 320 words in 1x16bit mode or 640 words in 2x8bit mode.
Using a methodology analogous to \cite{TC}, we implement the BC architecture as a full-custom design and performed HSpice energy and timing characterization. 
Targeting a 28nm CMOS technology from TSMC, the architecture can operate at a maximum frequency of 2.2GHz. The subarray requires 376pJ for reading a 16-bit word, and 414pJ energy for writing it.
An in-memory shift-add operation requires 381pJ. 
The subarray has an area of 1240$\mu m^2$, of which 26.5\%  is used for the BCU and the LGP circuits. 
6.5\% of the total area is used to implement the negation, embedded shift, and wordline parallelism.
Finally, 67\% of the area is filled up by the SRAM cells.

The GCW decoder is designed as semi-custom IC, as shown in \figref{fig:layout}(b). It is synthesized, placed and routed (again, in 28nm CMOS technology from TSMC) to extract its area, timing, and energy requirements. The circuit has an area of 760$\mu m^2$, which represents 61\% of the area of a single subarray, and less than 1\% of the total area in a 128 subarrays configurations. The decoder requires 1fJ per cycle to operate. 

We compare performances with the subarray design described in \cite{TC}. Such architecture has been shown to achieve 3x better performance, and 1.5x increased energy efficiency compared with the ARM NEON SIMD accelerator when running inferences on benchmark CNNs.

\section{Results}\label{sec:res}

\subsection{Accuracy-constrained compression}

\begin{figure}
    \centering
    \includegraphics[width = 0.8\linewidth]{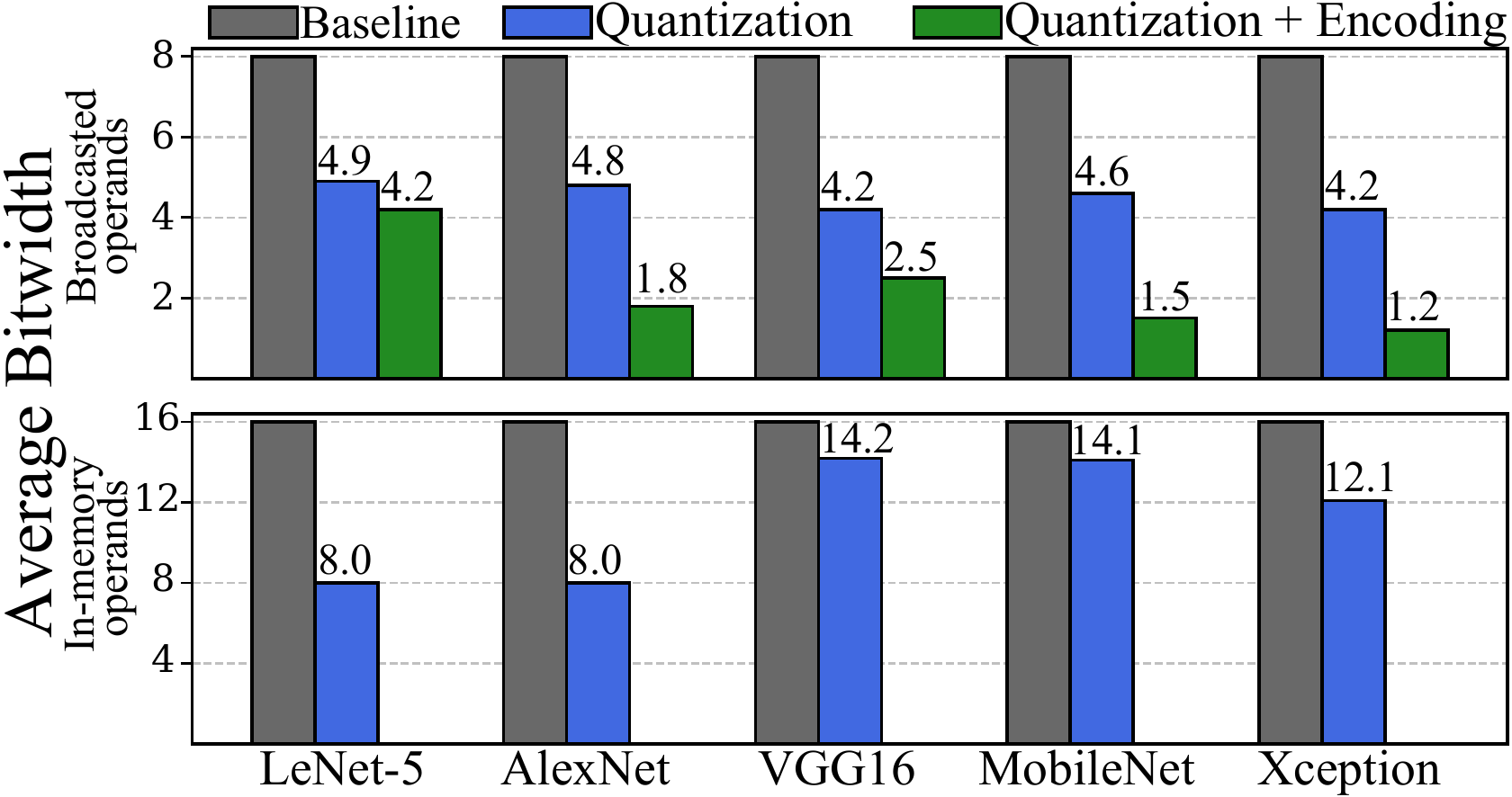}
    \caption{Average bitwidths achieved in our evaluated benchmarks by employing a synergic use of heterogeneous quantization (blue bars) and GCW encoding (green bars), for maximum accuracy degradation constraint of 1\%.}
    \label{fig:overall_compression}
    \vspace{-0.3cm}
\end{figure}

\figref{fig:overall_compression} depicts the average bitwidth (across layers) of IMOs and BOs in CNN benchmarks applications optimized with our proposed methodology. In all cases, results are for an accuracy threshold of 1\% with respect to implementations having 16-bits IMOs and 8-bits BOs. 

Blue bars illustrate the average bitwidth reduction achieved in convolutional and fully connected layers by mean of our heterogeneous quantization approach (as detailed in \secref{sec:mixed_precision}), which results in compression ratios of CNN models of 76.8\% on average.

Then, additional savings are achieved by encoding the weights of convolutional layers (illustrated in \secref{sec:hu}). Results are shown as green bars in \figref{fig:overall_compression}. They show that GCW encoding effectively contributes to the reduction of   storage requirements, resulting in overall model size savings of 85.3\% on average. 

Experimental outcomes show that all activations of convolutional layers of simpler CNNs (LeNet-5 and AlexNet) can be effectively reduced to 8-bits while abiding by the accuracy constraint. Such optimization instead can only be selectively applied in more complex benchmarks such as in VGG16, MobileNet and Xception, highlighting the benefit of a heterogeneous approach.

Moreover, especially high  compression ratios are achieved for larger models, because their size is largely determined by the footprint of convolutional weights. These can be effectively compressed by quantization and encoding (by up to more than 20x for the Xception CNN).

\subsection{CNN Inference Cycle-count Reduction}

\begin{figure}
    \centering
    \vspace{-0.2cm} 
    \includegraphics[width = 0.85\linewidth]{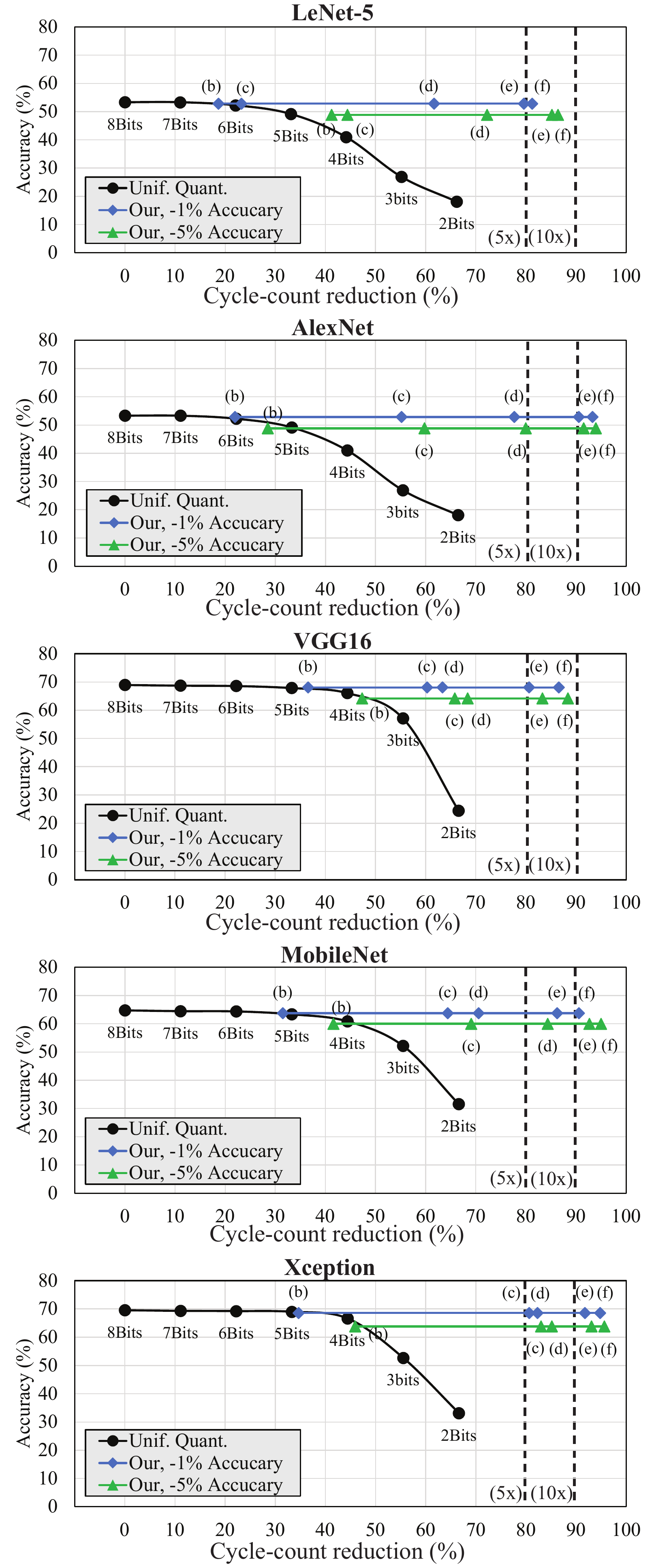}
    \vspace{-0.2cm} 
    \caption{Accuracy and cycle-count reductions in homogeneously quantized CNNs (black lines) and optimized CNNs (blue and green lines) for a 1\% and a 5\% accuracy drops. Data refers to single-subarray BC architectures. Vertical dashed lines mark speed-up levels of 5x and 10x.  }
    \vspace{-0.2cm} 
    \label{fig:cyclered}
\end{figure}

\begin{table*}
    \centering
    \caption{Inferences-per-second (left) and inference energy (right) in homogeneously quantized implementations executing on the BC architecture in \cite{TC} compared to  heterogeneously quantized models executed on optimized BC architecture and employing 1, 32 or 128 subarrays. Optimized implementations are obtained for a 1\% accuracy degradation threshold.}

    \includegraphics[width = 0.9\linewidth]{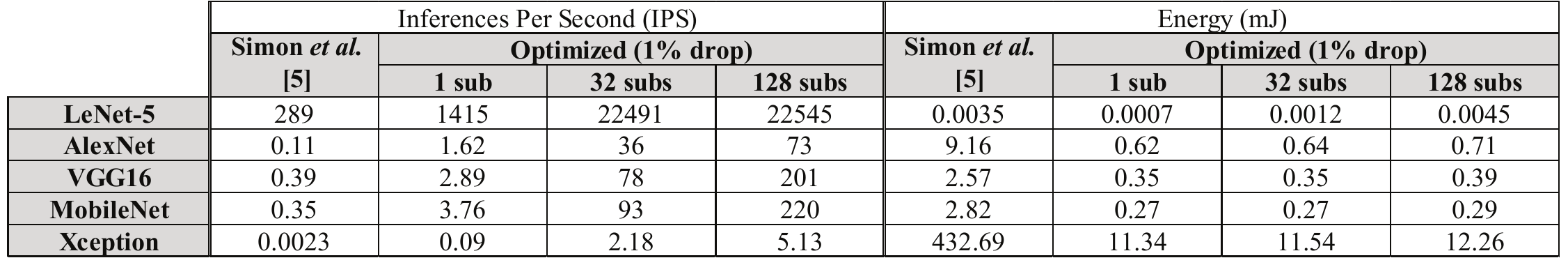}
    \label{fig:energyandperf}

\end{table*}

\begin{figure}
    \centering
    \vspace{-0.2cm} 
    \includegraphics[width = \linewidth]{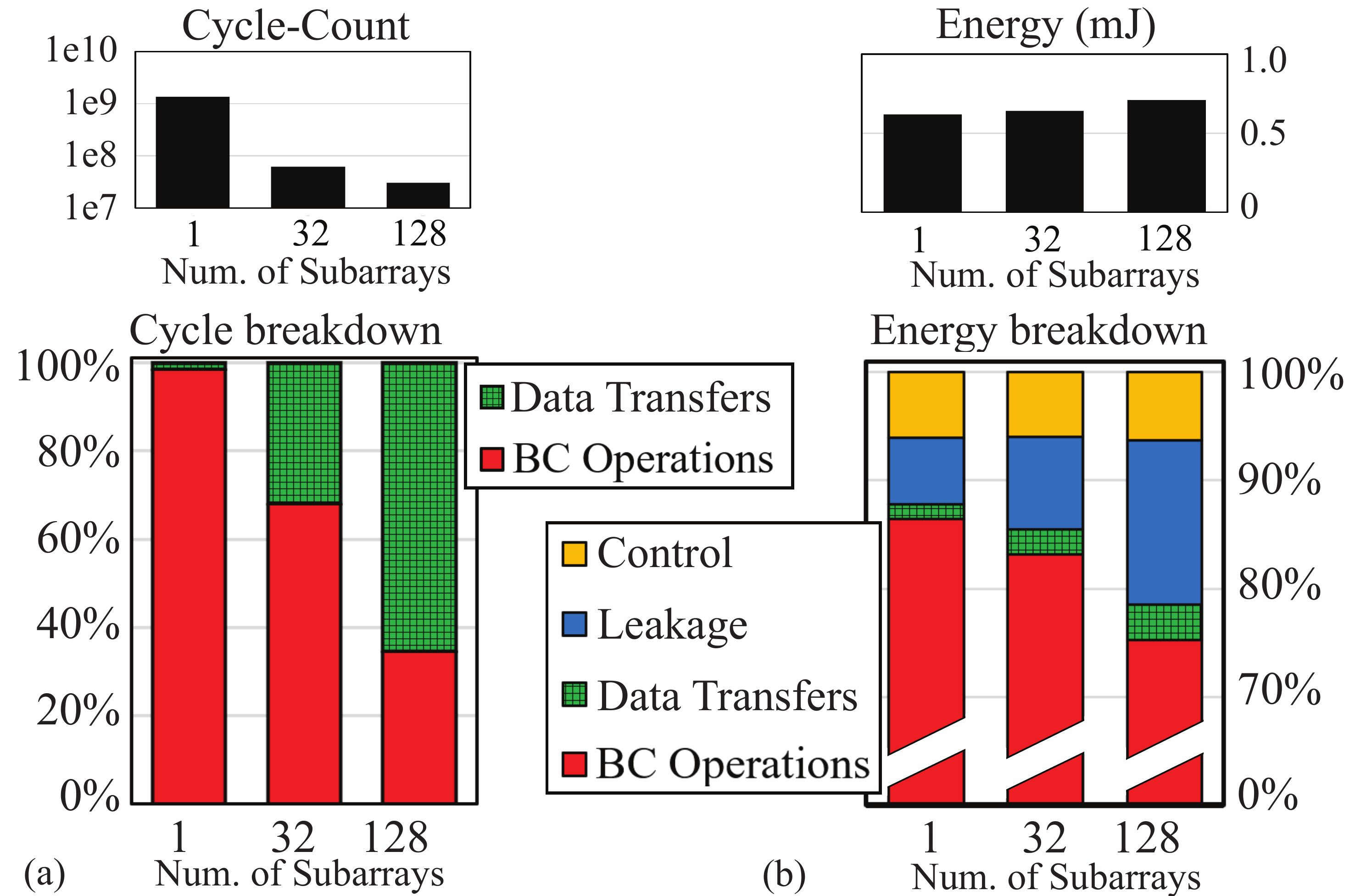}
    \vspace{-0.2cm} 
    \caption{(a) Cycle and (b) energy breakdown of multiple subarray architecture for an AlexNet inference with SW-HW optimization.}
    \label{fig:alexbreakdown}
    \vspace{-0.2cm} 
\end{figure}

\figref{fig:cyclered} illustrates the accuracy/performance trade-off in different optimized benchmarks. 
Black markers report the accuracy of homogeneously quantized models, while blue and green lines show the accuracy achieved in the various steps of the proposed hardware/software co-design methodology, for accuracy degradation thresholds of 1\% and 5\%, respectively, compared to configurations with 8-bit BOs and 16-bits IMOs.

In \figref{fig:cyclered}, the points (b), (c) and (d) highlight improvements obtained in the different stages of the optimization strategy, as illustrated in \figref{fig:flow} and detailed in \secref{sec:sw_opt}. They report performance/accuracy of CNNs after (b) BOs, (c) convolutional filters, and (d) IMOs optimization.

Points (e) and (f) report further cycle-count reductions obtained by hardware optimizations. In (e) up to three single-cycle bit-shifts (NES = 3 in Section \ref{sec:bc4}) are supported. Additionally, in (f), MAC operations involving zero-valued broadcasted operands are skipped (\secref{sec:decompr}).

When employing 5 bits, baseline uniform quantization achieves a 33\% cycle-count reduction at the cost of an average 1.3\% accuracy degradation, which rapidly increases for smaller bitwidths. Conversely, significantly higher performance improvements are obtained with our approach. In particular,  heterogeneous quantization alone (steps (b)-(d)) enables alone up to 80\% cycle-count gains. BC hardware optimizations are also highly effective (steps (e) and (f)). Support for NES=3  results in an average cycle-count reduction of 2.1x, which increases to 2.9x when also skipping multiplications involving zero-valued broadcasted operands. Considering all software and hardware optimizations,  the co-design framework achieves an average cycle-count reduction of 89.3\% (a speed-up of 11.5x) for 8-bits quantized baselines for 1\% degradation thresholds, and 91.9\% average cycle-count reduction (a speed-up of 15x) for 5\% accuracy degradation.


Performance gains are linearly correlated to energy savings, as they derive from reductions in shift-adds. Illustrating this aspect,
\tblref{fig:energyandperf} shows the Inferences Per Second (IPS) and inference energy in 8-bits quantized CNN baselines and in optimized models. As an example, the baseline AlexNet requires a 9.16mJ to perform one inference. Our optimized hardware and software reduce this cost by 93\%, to just 0.62mJ,  when executing on a single subarray. 

\subsection{BC Architectures with Multiple Subarrays}

As shown in \tblref{fig:energyandperf}, the IPS of our evaluated benchmarks scales up with the number of subarrays, as the workload is effectively distributed using the CNN mapping strategy proposed in \secref{sec:cas}. For AlexNet, VGG16, MobileNet, and Xception, on average a speed-up of 58x is reached when employing 128 subarrays with respect to a single one. Being a smaller network, LeNet-5 is less amenable to parallelization, reaching in this setting a 15x speed-up.

The performance of our solution when varying the number of subarrays  are further detailed in \figref{fig:alexbreakdown}(a) for the Alexnet benchmark. In its top part, the figure reports absolute cycle-counts and energy requirements. Below, we show the proportional breakdowns.
These results indicate that, when the workload is entirely run on a single subarray, 99\% of the clock cycles are used to perform MAC operations. However, the use of 32 or 128 subarrays reduces this percentage to 68\% and 35\% (respectively) because data-transfers are performed sequentially, while the parallelism of BC operations grows linearly with the number of subarrays.

Then, the energy results included in \tblref{fig:energyandperf} show that the use of a high number of subarrays has only a small impact on energy efficiency (while greatly reducing run-times). For AlexNet, VGG16, MobileNet, and Xception the difference in energy between 1 and 128 subarray configurations is only 12\% (again, LeNet-5 is an outlier due to its small dimensions).   Indeed, the number of subarrays does neither influence the energy cost of BC operations nor the number of BC operations required by an inference.
The slight increase in energy consumption is due to leakage energy (\figref{fig:alexbreakdown}(b)) consumed by subarrays during data transfers, and by the larger and more energy-hungry H-tree required to connect them. On the other hand, a higher level of parallelism requires decoding a smaller number of BC instructions, decreasing the related energy budget, which nonetheless accounts for less than 1\% of the total energy in all configurations.

\section{Conclusion} \label{sec:con}
Edge AI requires the execution of extremely computational and memory-intensive applications on constrained platforms. Bit-Line computing is a promising avenue to cope with this challenge, but demands careful synergic co-optimization of applications and hardware.
To tackle this problem, in this work we have presented a framework that comprises hardware-aware application optimizations as well as novel architectural solutions to effectively harness them. On the application side, CNNs are compressed by combining heterogeneous quantization and weights encoding. In turn, the proposed Bit-line Computing (BC) platform embeds low-overhead hardware features to perform run-time decoding, support fine-grained quantization of broadcasted operands, and leverage word-level parallelism of in-memory operands.
Results on a variety of CNN benchmarks have demonstrated that, for a 1\% accuracy degradation constraint, our compression strategy offers 85\% average memory reductions compared to uniformly quantized CNN implementations. Using the same constraint,  our proposed BC architecture achieves on average 11x inference speed-ups and 90\% energy savings with respect to state-of-the-art BC  approaches.

\ifCLASSOPTIONcompsoc
  \section*{Acknowledgments}
\else
  \section*{Acknowledgment}
\fi

This
work has been supported by 
the EC H2020 WiPLASH (Ga. No. 863337),  
the EC H2020 FVLLMONTI (Ga. No. 101016776),
the ERC Consolidator Grant COMPUSAPIEN (Ga. No. 725657),
and by the Swiss NSF ML-Edge (Ga. No. 182009)
projects.

\ifCLASSOPTIONcaptionsoff
  \newpage
\fi

\bibliographystyle{IEEEtran}
\bibliography{biblio.bib}

\begin{IEEEbiography}[{\includegraphics[width=1in,height=1.25in,clip,keepaspectratio]{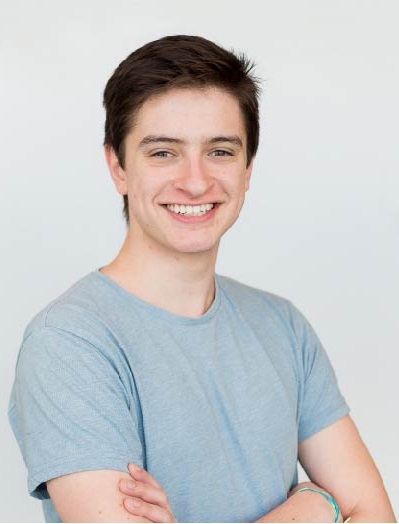}}]{Marco Rios}
received the M.Sc. degree in Computer Science and Electronics
For Embedded Systems from Université Grenoble Alpes, France, in 2018. He is currently a PhD student at the Embedded Systems Laboratory of EPFL, Switzerland. His research interests include design of integrated systems and circuits, in-SRAM computing and the system impact of emerging memories.
\end{IEEEbiography}

\begin{IEEEbiography}[{\includegraphics[width=1in,height=1.25in,clip,keepaspectratio]{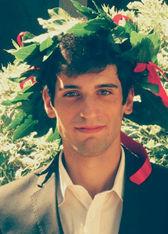}}]{Flavio Ponzina}
received the M.Sc. degree in Computer Engineering from Politecnico di Torino, Italy, in 2018. He is currently a PhD student at the Embedded Systems Laboratory (ESL), EPFL. His main research interests include low power architectures and AI-based systems optimization.
\end{IEEEbiography}

\begin{IEEEbiography}[{\includegraphics[width=1in,height=1.25in,clip,keepaspectratio]{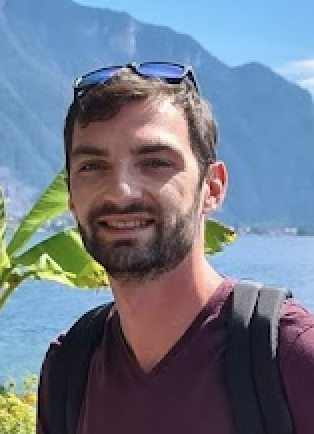}}]{Alexandre Levisse}
received his Ph.D. degree in Electrical Engineering from CEA-LETI, France, and from Aix-Marseille University, France, in 2017. From 2018 to 2021, he was a post-doctoral researcher in the Embedded Systems Laboratory at the Swiss Federal Institute of Technology Lausanne (EPFL). From 2021, he works as a scientist in EPFL. His research interests include circuits and architectures for emerging memory and transistor technologies as well as in-memory computing and accelerators.\end{IEEEbiography}

\begin{IEEEbiography}[{\includegraphics[width=1in,height=1.25in,clip,keepaspectratio]{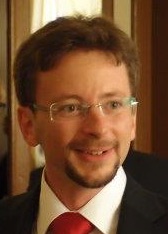}}]{Giovanni Ansaloni}
is a researcher at the Embedded Systems Laboratory of EPFL (ESL-EPFL, Lausanne, CH). He previously worked as a Post-Doc at the University of Lugano (USI, CH) between 2015 and 2020, and at EPFL between 2011 and 2015. He received a Ph.D. degree in Informatics from USI in 2011. His research efforts focus on domain-specific and ultra-low-power architectures and algorithms for edge computing systems, including hardware and software optimization techniques.\end{IEEEbiography}

\begin{IEEEbiography}[{\includegraphics[width=1in,height=1.25in,clip,keepaspectratio]{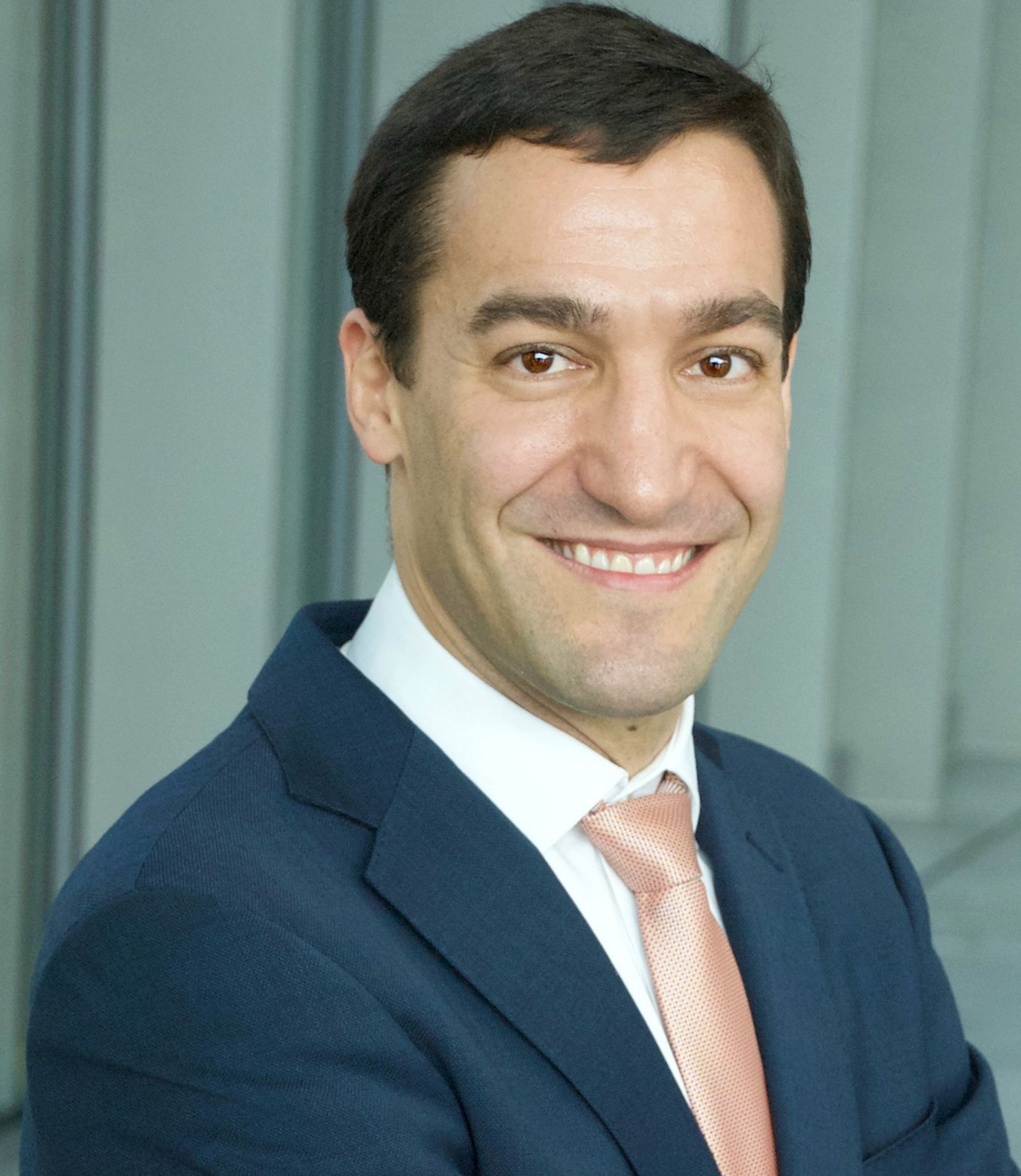}}]{David Atienza}
is a full professor of electrical and computer engineering, and head of the Embedded Systems Laboratory (ESL) at EPFL, Switzerland. He received his Ph.D. in computer science and engineering from UCM, Spain, and IMEC, Belgium, in 2005. His research interests include system-level design methodologies for multi-processor system-on-chip (MPSoC) servers and edge AI architectures. He has co-authored more than 350 papers, one book, and 12 patents. Dr. Atienza has received, among other recognitions, the ICCAD 10-Year Retrospective Most Influential Paper Award in 2020, the Most Influential DAC Under-40 Innovators Award in 2018, and an ERC Consolidator Grant in 2016. He is an IEEE Fellow and an ACM Distinguished Member.\end{IEEEbiography}

\end{document}